\documentclass[10pt, letterpaper]{article}

\title{A synopsis of comparative metrics for classifications}
\author{Bernardo Lopo Tavares\footnote{Instituto Superior T\'ecnico - Universidade de Lisboa. This article is part of the author's MSc thesis, under coordenation of Prof. Alexandre Francisco and Prof. Pedro Martins Rodrigues.}\\ \texttt{bernardo.lopo@ist.utl.pt}}

\date{April 2018}

\usepackage{textcomp}
\usepackage{cite}
\usepackage[utf8]{inputenc}
\usepackage{amsthm}
\usepackage{amssymb}
\usepackage{amsmath}
\usepackage{graphicx}
\usepackage{multirow}
\usepackage{float}
\usepackage[normalem]{ulem}
\usepackage[nottoc]{tocbibind}
\newtheorem{dfn}{Definition}[section]

\newtheorem{nota}{Notation}[section]
\newtheorem{thm}{Theorem}[section]
\newtheorem{ex}{Example}[section]
\newtheorem{rmk}{Remark}
\newtheorem{alg}{Algorithm}

\begin{document}

\maketitle

%%%The work in this chapter is mainly expositive and a lifting from a collection of papers and articles. You can find the references on the appropriate section.

\begin{abstract}
Phylogeny is the study of the relations between biological entities. From it, the need to compare tree-like graphs has risen and several metrics were established and researched, but since there is no definitive way to compare them, its discussion is still open nowadays. All of them emphasize different features of the structures and, of course, the efficiency of these computations also varies. The work in this article is mainly expositive (a lifting from a collection of papers and articles) with special care in its presentation (trying to mathematically formalize what wasn't presented that way previously) and filling (with original work) where information wasn't available (or at least, to our knowledge) given the frame we set to fit these metrics, which was to state their discriminative power and time complexity. The \textit{Robinson Foulds}, \textit{Robinson Foulds Length}, \textit{Quartet}, \textit{Triplet}, \textit{Triplet Length}, \textit{Geodesic} metrics are approached with greater detail (stating also some of its problems in formulation and discussing its intricacies) but the reader can also expect that less used (but not necessarily less important or less promising) metrics will be covered, which are \textit{Maximum Aggreement Subtree}, \textit{Align}, \textit{Cophenetic Correlation Coeficcient}, \textit{Node}, \textit{Similarity Based on Probability}, \textit{Hybridization Number} and \textit{Subtree Prune and Regraft}. Finally, some challenges that sprouted from making this synopsys are presented as a possible subject of study and research. 
\end{abstract}

%%%%%%%%%%%%%%%%%%%%%%%%%%%%%%%%%%%%%%%%%%%%%%%%%%%%%%%%%%%%%%%%%%%%%%%%
\section{Problem Statement}
\label{section:ooverview}

The topic of this work is comparisson metrics. Given two classifications, the challenge is to generate some parameter that expresses how similar these classifications are.
A \textbf{classification} is some \textit{data structure} that expresses relation between the data. For instance:

One of the big applications of comparisson metrics is \textbf{phylogenetics}. \textbf{Phylogenetics} is the study of relations between biological entities: may it be genes, species or individuals. If we consider Darwin's theory of evolution, we can think of every species as having a direct ancestor (or multiple direct ancestors, but for sake of this example, lets assume there's only one) and multiple sucessors, and then, it's only natural to suggest that the whole map of species evolution can be described by a \textit{tree-like graph}. We say that one specific tree hypothesis for the arrangement of species is a \textbf{classification}.

One of the first \textit{data structures} considered for the first known problem of classification were the dendrograms. \textbf{Dendrograms} are tree graphs where the data is concentrated in a specific part of the graph, while the rest expresses the relation between it. This was a way to relate data by hierarchical clustering: close data is on the same cluster (cluster as a \textit{bulk} of close related data), and how closer they are on the tree dictates how alike the information stored in the leaves is. The first effective numerical method to compare dendrograms was developed by Sokal and Rohlf in 1962, is known as the \textbf{cophenetic correlation} and it will be defined in the appropriate section. However, at the time, it was a method created to compare dendrograms generated from numerical taxonomic research.

However, since these methods were born out of a necessity to compare \textit{tree-like structures}, a lot of other metrics were proposed, some of them formalised for quirky data structures and to emphasize different properties (based on topology, edge weight and other parameters). This happened since there was no definitive method to compare trees, from a discriminative point of view and also from lack of efficiency: in computer science, graphs are a complicated structure to work with, and with the growth of the field of application and amount of available data overtime, it's no surprise the need for an efficient and personalized way to deal with these problems rises.

In recent years this field of study has widened and our knowledge of this problem deepened, since more people started working on it. Probably the most relevant \textbf{metric} (or \textbf{distance}) is \textit{Robinson Foulds} since it can generate a parameter in linear time in the number of vertices of our trees (which is fairly good), but, as stated before, this doesn't mean that this metric suits all problems, hence the need to create other metrics. In this article our main goal is to go over these metrics, formalizing them and discussing its relevant aspects such as the advantages, disadvantages and what differences it from the others.

%%% chapter our main goal is to go over these metrics, formalizing them and discussing its relevant aspects such as the advantages, disadvantages and what differences it from the others.

\raggedbottom

\section{Methodology and relevant aspects}
\label{section:methodology}

As stated before, no metric should be considered as default for all problems. Depending on the problem at hand, choosing a metric over the other can be an advantage given the goal we want to achieve.
However, this idea revolves around two important concepts. Since graphs are dificult to handle from a computation point of view, it's an advantage to know how fast are we able to compute the metric. On the other hand, since the output parameter describes how close two trees are the metric might benefit certain properties over others. These two can be referred as \textbf{efficiency} and \textbf{discriminatory power}, respectively.

\textbf{Efficiency} can be seen from a \textbf{complexity} point of view, but complexity varies with the implementation of the algorithms. Althrough most of the times the description for the metrics might describe an algorithm, it might exist an equivalent algorithm implementation that is not the literal \textit{translation} from that description but outputs the same values for the same inputs, althrough has a lower complexity.

The \textbf{Discriminatory Power} will depend on the metric. For instance, some metrics might benefit the tree topology over edge length (or weight) and some might have associated errors or output the same value for some types of trees. Beeing aware of these properties it's important to recognise the best metric to solve a problem at hand, however (and unlike efficiency) it's not a parameter that we can quantify, so we will go over the discrimatory power of each metric on the apropriate section.

The following work is structured according to the analysed metrics. First we will go over the most commonly used metrics and then the others. The latter have a less common use since they may be formalized for a different type of data structures (e.g.:, \textit{Hybridization number}) or aren't that practical to compute in present days (e.g.: \textit{Subtree Prune and Regraft}). In each subsection our main focus (besides defining the metrics) will be their \textbf{efficiency} and \textbf{discriminatory power}.

\section{Basic definitions}
\label{section:basicdef}

Next we'll go over some definitions needed to define the metrics later on.

\begin{dfn}{\textbf{Graph theory basic definitions}} \\
Let $G=(V,E)$ where $V$ is a set of vertices and $E$ a set of edges. An edge is a pair $(v_1,v_2)$ of vertices from $V$ (for a lighter notation, one may write $v_1v_2$ instead of $(v_1,v_2)$). $G$ is a \textbf{graph}.\par
A \textbf{path} is a subset $P\subseteq E$ of size $k$ that can be ordered in a way that for the $i$-th element of $P$ $(v_{i,1},v_{i,2})$: $v_{i,1}=v_{i-1,2}$, $v_{i,2}=v_{i+1,1}$. We say that a path is a \textbf{cycle} if, on top of beeing a \textit{path}, $v_{0,1}=v_{k,2}$. We name a graph without cycles a \textbf{tree graph} (or just \textbf{tree}). The length of a \textit{path} $P$ with no cycles is $|P|$.\par
In \textbf{undirected graphs} the edges $(a,b)$ and $(b,a)$ are equal. We will work with undirected graphs unless is differently stated\par
Let $u,v\in V$. We say that $v$ is \textbf{neighboor} of $u$ if $(u,v)\in E$ and we write $u\sim v$. In a \textit{undirected graph} this relation is reflexive. The \textbf{degree} of a vertex is the number of neighboors it has.
\end{dfn}

For the next set of concepts, it's important we assume that graph vertices can have \textbf{labels}, this means that exists a function $label:V\longrightarrow \mathrm{STR}\cup\{'\mathrm{NULL}'\}$ that for each vertex returns a \textit{string} (that we call \textit{label}) or $\mathrm{NULL}$. The same way we define a concept of edge \textbf{weight} (or \textbf{length}), as a function $weight:E\longrightarrow\mathbb{R}$. That's actually needed to be considered for some problems and could represent, building over the \textit{phylogeny} application, for example, how many years are between species. In both cases they are just ways to hold information in these data structures, if needed. 

\begin{dfn}{\textbf{Tree specific concepts}}\\
Let $T=(V,E)$ be a \textit{tree graph}.
\begin{itemize}
  \item A \textbf{leaf} is a vertex with degree $1$.
  \item If exists one (and only one) vertex $v\in V$ such that $label(v)='\mathrm{root}'$ then we say that $v$ is the \textbf{root vertex} and that $T$ is a \textbf{rooted tree}. If a \textit{root vertex} doesn't exist, then $T$ is an \textbf{unrooted tree}. Assuming $T$ is \textit{rooted} we can now define a new set of concepts:
      \begin{itemize}
        \item The \textbf{depth of a node} $v$ is the number of edges on the path (that on a tree is singular) from the \textit{root} vertex to $v$. The \textbf{depth of a tree} is the maximum depth between all nodes. (We will refer to the depth of a vertex $v$ as $depth(v)$, and the depth of a tree $T$ as $depth(T)$)
       %%% \item We say that $T$ is a \textbf{clocklike tree} if the leaves of $T$ have different depths.
        \item Let $u, v\in V$. We say that $u$ is an \textbf{direct ancestor} of $v$ if $(u,v)\in E$ and $depth(u)<depth(v)$. In this case, $v$ is also a \textbf{direct descendant} of $u$. Also, we generally say that $u$ is an \textbf{ancestor} of $v$ if there is a path of direct ancestors from $v$ to $u$ (similarly we define the same general concept for \textbf{descendant}).
        \item A \textbf{clade} consists of a vertex and all its lineal descendants.
      \end{itemize}

  \item A \textbf{dendrogram} is a tree where only leaves (and, in case of a rooted tree, the root) have labels.

  \item A \textbf{forest} is a colection $F=\{T_1,T_2,...,T_k\}$ where for every $i$ $T_i$ is a tree.

\end{itemize}
\end{dfn}

\begin{dfn} 

Let $d:X\times X \longrightarrow \mathbb{R}^+_0$ be an injective function. We say that $d$ is a \textbf{metric} over $X$ (and $d$ is called the \textbf{distance function}) if: \textbf{(1)} $d$ is non-negative; \textbf{(2)} $d(x,y)=0$ if and only if $x=y$; \textbf{(3)} $d$ is symmetrical, that is $d(x,y)=d(y,x)$; \textbf{(4)} $d$ satisfies the triangular inequality, that is for all $x,y,z\in X$ $d(x,z)\leq d(x,y)+d(y,z)$.

\end{dfn}

In regard to efficiency, we now define a notation that will be usefull to talk about program complexity.

\begin{dfn}{\textit{Big-Oh Notation}}\\
If $f$ and $g$ are two functions from $\mathbb{N}$ to $\mathbb{N}$, then we: \textbf{(1)} say that $f=O(g)$ if there exists a constant $c$ such that $f(n)\leq c \cdot g(n)$ for every sufficiently large $n$, \textbf{(2)} say that $f=\Omega(g)$ if $g=O(f)$, \textbf{(3)} say that $f=\Theta(g)$ if $f=O(g)$ and $g=O(f)$, \textbf{(4)} say that $f=o(g)$ if for every $\epsilon > 0$, $f(n)\leq \epsilon \cdot g(n)$ for every sufficiently large $n$, and \textbf{(5)} say that $f=\omega(g)$ if $g=o(f)$.\\
To emphasize the input parameter, we often write $f(n) = O(g(n))$ instead of $f = O(g)$, and use similar notation for $o$, $\Omega$, $\omega$, $\Theta$.
\end{dfn}

When one refers \textit{complexity} it's common to use \textit{Big-Oh Notation}, but there are other notations. This will be important to understand how the computation time varies with the size of the input. So if we say that an implementation runs in $O(n)$ time it means time grows linearly with input growth. As expected, \textit{how fast} a program runs will depend on its complexity. Given two functions $f$ and $g$, $f=O(p_f(n))$ and $g=O(p_g(n))$ the function with higher complexity is determined by which of $p_f(n)$ and $p_g(n)$ as a higher growth rate. For example, if $p_f(n)=e^n$ and $p_g(n)=n^5+n^3+10$, then $f$ has a higher complexity.

%%%%%%%%%%%%%%%%%%%%%%%%%%%%%%%%%%%%%%%%%%%%%%%%%%%%%%%%%%%%%%%%%%%%%%%%
%%%%%%%%%%%%%%%%%%%%%%%%%%%%%%%%%%%%%%%%%%%%%%%%%%%%%%%%%%%%%%%%%%%%%%%%
%    Usual Approaches to the problem                                   %
%         Usual distances used to approach the problem                 %
%           - Robinson Foulds, Robinson Foulds Length                  %
%           - Triplets, Triplets Length, Quartet                       %
%           - Geodesic distance                                        %
%%%%%%%%%%%%%%%%%%%%%%%%%%%%%%%%%%%%%%%%%%%%%%%%%%%%%%%%%%%%%%%%%%%%%%%%

\section{Usual Approaches}
\label{section:uapproach}

In this section the reader will find the most used metrics in comparing classifications. Reading this first subsection about \textit{Robinson Foulds} and \textit{Robinson Foulds Length} is advised even if not your point of interest since it might introduce concepts or notation that will be important later on for other metrics.

\subsection{Robinson Foulds, Robinson Foulds Length}
\label{subsection:rf}

Arguabily one of the most used metrics for comparing classifications, \textbf{Robinson Foulds} was result of the continued work of David F. Robinson and Leslie R. Foulds (published in 1981 on the Mathematical Biosciences journal) to compare \textit{phylogenetic trees}.

Originally, this metric was defined for \textit{binary dendrograms}, however, it can be used without adaptation for most tree structres.

\begin{dfn} \label{dfn:rf1} {\textbf{Robinson-Foulds distance}}\\
Let $A$ and $B$ be two rooted trees with the same number of leaves and $C_A$ and $C_B$ the set of all clades for $A$ and $B$ respectively. The \textbf{Robinson-Foulds} distance $d_{RF}$ is defined as $$d_{RF}(A,B)=|C_A\backslash C_B|+|C_B\backslash C_A|$$
\end{dfn}

There's obviously an abuse of notation on the previous definition, since $C_A$ and $C_B$ aren't built over the same set of vertices and edges (but instead over $V_A$, $V_B$ and $E_A, E_B$ respectively), but that won't be a problem since we won't work with this definition, this should only give the reader an abstract idea of how the metric works. However, this was not the first definition where they settled. Actually, the original motivation behind this later formalized definition comes from the attempt to know how far was, given two trees, one from the other, considering a specific operation that consisted in \textit{glueing} adjacent vertices and erasing the edge between or (for the inverse operation) splitting one vertex in two new vertices connected by a new edge.

To talk about the background of the \textit{Robinson Foulds distance} we need some extra notation:

\begin{nota} Given a tree $T=(V,E)$ we define $S=\{x\in \mathrm{STR}: \exists v\in V\ s.t.\ label(v)=x\}$ (that is $S=label(V)$) as the \textbf{set of labels} of $T$.
A \textbf{labeled tree} consists of a \textit{3-tuple} $T_l=(V,E,S)$ where $(V,E)$ is a tree and $S$ the corresponding set of labels.
The set of all labelled trees with $S$ as the set of labels is denoted by $\gamma_S$; We say that two trees are \textbf{identical} if there is a bijective map between them that preserves labeling, meaning that, for two identical labelled trees $A,B\in\gamma_S$ exists $h:V_A\longrightarrow V_B$ bijective, such that $x,y\in V_A$ and $xy\in E_A$ if and only if $h(x)h(y)\in E_B$ and $label(x)=label(h(x))\wedge label(y)=label(h(y))$.
\end{nota}

\begin{rmk}
In all extension of our work we assume that $S$ is the set of labels of the leaves (and usually consists in all natural numbers until some $k\in\mathbb{N}$), meaning that leaves of trees in $\gamma_S$ must be exactly $|S|$ and its labels will be non repeated labels from $S$. The label 'root' is not in $S$. 
\end{rmk}

\begin{dfn}
(\textbf{Operation} $\alpha$)\\
Let $T=(V,E,S)\in\gamma_S$, $|V|=m$ and $v_iv_j\in E$. Then, $\alpha : \gamma_S \times E \longrightarrow \gamma_S$ is a function such that $\alpha(T,v_i v_j)=(V',E',S)$ where:
\begin{itemize}
  \item $V'=(V\cup\{v_{m+1}\})\backslash\{v_i,v_j\}$;
  \item $E'=[(E\backslash E^i)\backslash E^j]\cup \{v_{m+1}v_h: v_hv_i\in E^i\ or\ v_hv_j\in E^j, h\neq i\ or\ h\neq j\}$, where $E^k=\{v_kv_q:v_kv_q\in E, v_q\in V\}$ is the set of edges incident with $v_k$.
  \item $label(v_{m+1})=label(v_i)\cup label(v_j)$
\end{itemize}
\end{dfn}

The reader should understand that this operation does nothing more than to collapsing edges and vertices on their ends into a new vertex $v_{m+1}$. One should note as well that, in this case, the co-domain of the $label$ function is the \textit{powerset} of $STR$, or information on the labels would be lost between $\alpha$ operations.

We spare ourselves the work of formalizing a definition for the operation $\alpha^{-1}$ but one should have a straightforward idea of how it works, beeing less straightforward only in regards of label and neighborhood distribution between the new vertices. Since the objective is to transform, with applying the minimum amout of $\alpha$ and $\alpha^{-1}$ operations, one tree into another, one should choose the distribution of neighboors and labels according to whats best to reach that end, when applying $\alpha^{-1}$.

The operations $\alpha$ and $\alpha^{-1}$ are also called as \textbf{contraction} and \textbf{decontraction of Bourque} respectively.

This leaves us with the original definition stated:

\begin{dfn}{\textbf{Robinson-Foulds distance}}\\
Let $S$ be a set of labels and $A,B\in\gamma_S$. The Robinson-Foulds distance between $A$ and $B$, $d'_{RF}(A,B)$, is defined as the minimum number of contractions and decontractions of Bourque necessary to apply on $A$ to get $B$.
\end{dfn}

One should note that, given three rooted trees $A,B,C\in\gamma_S$:

\begin{itemize}
  \item $d'_{RF}(A,B)=0$ then $A$ is identical to $B$;
  \item $d'_{RF}(A,B)>0$ then $A$ is not identical to $B$;
  \item $d'_{RF}(A,B)=d'_{RF}(B,A)$;
  \item $d'_{RF}(A,C)\leq d'_{RF}(A,B) + d'_{RF}(B,C)$.
\end{itemize}

All these items should cause no trouble for the reader to prove as true considering the definition so, in fact, $d'_{RF}$ is a well defined metric. If you are looking for the proof that $d'_{RF}$ is actually the same as $d_{RF}$ from Definition \ref{dfn:rf1} you should look into \cite{bib:10}, taking in consideration that in that article the conclusion is reached not in terms of \textit{clades} but of \textit{edges}, but they are actually the same since there is a one-to-one correspondence between edges and clades in these structures (removing an edge from a tree would lead to an unconnected graph with two connected components. The one-to-one correspondence is given by associating that edge with the connected component that contains the deepest vertex of the two vertices the edge was connecting). The main idea to understand the equality lies on the existence of a third \textit{mid way} tree $C\in\gamma_S$ between the sequence of applying the $\alpha$ and $\alpha^{-1}$ operations that contains the clades that are both in $A$ and $B$ (for $A,B\in\gamma_S$), and one should account $1$ for each collapsed and generated edge in this process. We'll talk about this \textit{mid way} tree later (Definition \ref{dfn:scmsct}).

When it comes to implementation of the algorithm to calculate this metric, most authors refers it as fairly simple, but wasn't until William H. E. Day formalized an algorithm in 1985 that showed that to compute this was actually a linear time problem.

Given the result reached in \cite{bib:10}, the problem shifted from counting the number of $\alpha$ and $\alpha^{-1}$ operations between the two trees (which could be seen as an actual challenge to compute) to counting clades. In William H. E. Day article \cite{bib:11} he actually solves the problem for a group of similar problems in the field of study, which include the implementation of the Robinson Foulds distance as well. With little expression manipulation one can conclude that $d_{RF}$ is actually also equal to, given two trees $A$ and $B$
$$d_{RF}(A,B)=|C_A\backslash C_B|+|C_B\backslash C_A|=|C_A|+|C_B|-2|C_A \cap C_B|$$
where $C_X$ is the set of clades of the tree $X$. Since the number of clades of a generic tree is easily calculated in linear time (given the one-to-one correspondence we aproached earlier), the problem is reducted to calculate the clades of $A$ that are also clades of $B$.

However, Day refers to \textit{clades} indirectly, since he works with \textbf{clusters} through the whole article, which can be seen as the sets of labels on the clade's leaves. Therefore we can also formalize a \textbf{cluster representation} for trees.

\begin{nota}
\label{nota:day}
  In William Day's work, for every tree $T$ there is a cluster representation given by a set of sets of labels. Each set of labels is in fact the set of labels of the leaves for every clade of the tree. This cluster representation is denoted as $T'$.

\end{nota}

To calcule the $d_{RF}$ distance, Day actually formalizes a new structure which he calls as \textit{Strict Consensus Tree}:

\begin{dfn} \label{dfn:scmsct} (\textbf{Strict Consensus Method}, William Day (1985))
Let $S$ be a set of labels and $C: (\gamma_S)^k \longrightarrow \gamma_S$ a function such that, for $T_1, T_2, ..., T_k \in \gamma_S$, $$(C(T_1,T_2,...,T_k))'=\bigcap_{1\leq i \leq k} (T_i)'$$
In this case, we say that $C$ is a \textbf{strict consensus method} and $C(T_1,T_2,...,T_k)$ the \textbf{strict consensus tree} between $T_1, T_2, ..., T_k$.
\end{dfn}

\textbf{Lets add to Day's definition that not only $C(T_1,T_2,...,T_n)$ has to satisfy its condition, but also it's the smaller tree (in terms of vertex count) to satisfy it.}

This means that, looking back on our Robinson Foulds distance, the $|C_A \cap C_B|$ term of the expression can be rewritten as $|(C(A,B))'|$. The algorithm defined in Day's paper is, in fact, an algorithm to calculate the strict consensus tree between $T_1,T_2,...,T_k\in\gamma_S$ with $|S|=n$ and the conclusion is that this algorithm is capable of doing it in $O(kn)$ time. Implementation, complexity reasoning and respective empirical verification is available in the article \cite{bib:11}.

Even though these results and proofs were published, as referred earlier, in 1981, David Robinson and Leslie Foulds gave us a blink of it in 1978 in \textit{Lecture notes in Mathematics, vol. 748} \cite{bib:12}. However, this 1978 text was actually a revision to a previously submited work that never got published: the unpublished work specified the \textit{Robinson Foulds} metric that we just discussed, and the published work specified that this unpublished metric was actually a particular case of a new metric there proposed, particular case in which every edge of the classifications feeded to this new distance had weight $1$. Later on we will understand that this was not the case, since $d_{RFL}$ has problems in its formulation.

\begin{nota}
A \textbf{weighted labeled tree} consists of a \textit{4-tuple} $T_{wl}=(V,E,S,w)$ where $(V,E,S)$ is a labeled tree and $w$ the corresponding \textit{weight function} $w:E\longrightarrow\mathbb{R}^+_0$.
The set of all weighted labelled trees with $w$ as weight function and $S$ as the set of labels is denoted by $\gamma^w_S$.  We say that two trees are \textbf{weight-identical} if there is a bijective map between them that preserves labeling and weight of the edges, meaning that, for two weight-identical trees $A,B\in\gamma^w_S$ exists $h:V_A\longrightarrow V_B$ bijective, such that $x,y\in V_A$ and $xy\in E_A$ if and only if $h(x)h(y)\in E_B$, $label(x)=label(h(x))\wedge label(y)=label(h(y))$ and $w(xy)=w(h(xy))$.
\end{nota}

Once we widen the \textit{Robinson Foulds} metric, we also realize that this new distance would need to account not only for the difference in tree topology but for edge's lengths as well. With this in mind, and not wanting to increse the complexity of the problem, \cite{bib:11} formalizes \textbf{Robinson Foulds Length} after presenting the following definitions:

\begin{dfn}\label{dfn:partfunc} {\textbf{Partitioning Function}}\\
Let $A\in\gamma^w_S$ such that $A=(V,E,S,w)$ and $Z_S$ the set of all proper partitions of $S$ into two subsets. Let $f:E\longrightarrow Z_S$ such that, for every edge $e\in E$, $f(e)$ returns the set of $Z_S$ that corresponds to the partition of $S$ given by (according to \textit{Day}'s notation, Notation \ref{nota:day}) the clusters of both connected components of $(V,E\backslash \{e\}, S, w)$. We say that $f$ is the \textbf{partitioning function} of $A$.
\end{dfn}

\begin{figure}[!htb]
  \centering
  \includegraphics[width=0.8\textwidth]{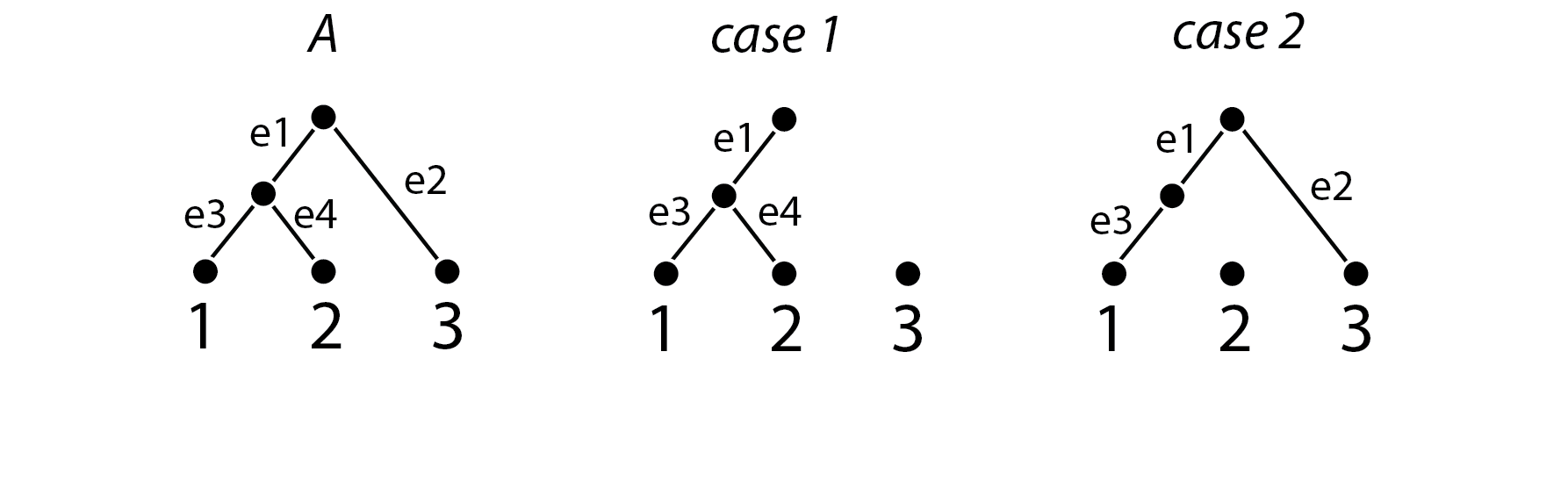}
  \caption[Caption for figure in TOC.]{ Depiction of the various connected components of the graph on the left (with leaf set $S=\{``1",``2",``3"\}$) uppon removal of $e_2$ and $e_4$.}
  \label{fig:PF}
\end{figure}

\begin{ex}
  Consider the tree $A\in \gamma^w_{\{``1",``2",`3"\}}$, depicted in Figure \ref{fig:PF}, and with partitioning function $f_A$. In case $1$, removing the edge $e_2$ would lead to the depicted connected components, concluding that $f_A(e_2)=\{\{``1",``2"\},\{``3"\}\}$. In case $2$, the same reasoning will lead us to conclude that $f_A(e_4)=\{\{``2"\},\{``1",$\\$``3"\}\}$.
\end{ex}

\begin{dfn} \label{dfn:medg}
  Let $A,B\in\gamma^w_S$ with edge sets $E_A$ and $E_B$ and partitioning functions $f_A$ and $f_B$. The edges $e_A\in A$ and $e_B\in B$ are \textbf{matched} if and only if $$f_A(e_A)=f_B(e_B)$$

\end{dfn}

This last definition can help us build concepts for matching functions from $E_1$ to $E_2$, and actually, one of those is needed for the definition of the \textit{Robinson-Foulds Length} distance. Consider that $h_{(A,B)}:E_A\longrightarrow E_B$ is a function that, given $e_A\in E_A$, if exists $e_B\in B$ such that $f_A(e_A)=f_B(e_B)$ then $h_{(A,B)}(e_A)$ is defined and equals $e_B$, undefined otherwise.

\begin{dfn}{\textbf{Robinson-Foulds Length distance}}\\
Let $A,B\in\gamma^w_S$, $E_A$, $E_B$, $E_{C(A,B)}$ the respective sets of edges, $f_A$ and $f_B$ the respective partitioning functions and $h_{(A,B)}$ the matching function from $A$ to $B$. The \textbf{Robinson-Foulds Length distance} is defined as

$$ d_{RFL}(A,B)= \Bigg(\sum _{e\in (E_A\backslash E'_A)} w(e) \Bigg) + \Bigg(\sum _{e\in (E_B\backslash E'_B)} w(e)\Bigg) +$$ $$+ \Bigg(\sum _{e\in E'_A} | w(e) - w(h_{(A,B)}(e))|\Bigg)$$

where:

$$ E'_A=\{e_A:e_A\in E_A, \exists e_B\in E_B\ s.t.\ f_A(e_A)=f_B(e_B) \}$$
$$ E'_B=\{e_B:e_B\in E_B, \exists e_A\in E_A\ s.t.\ f_B(e_B)=f_A(e_A) \}$$

\end{dfn}

%%%%\begin{rmk}
%%%%  \textbf{(comment)VERIFICAR A VERACIDADE E FAZER A PROVA; AFINAL ACHO QUE ISTO NÃO DÁ PARA FAZER ASSIM} If you are following \cite{bib:11} you may have noticed that the definition for $d_{RFL}$ is written with a set $E_1'$ that we didn't covered on our work. That set has a close relation to the set of edges of the strict consensus tree and for purposes of the definition it hasn't any implications or needs for modifications. It shouldn't be hard for the reader to understand and verify it.
%%%%\end{rmk}

After defining this new metric there's a few important things to note: first, how this distance behaves in a usual scenario considering its application, secondly, the relation between $d_{RF}$ and $d_{RFL}$, the relation with the sets $E'_A$ and $E'_B$ with the strict consensus tree of $A$ and $B$ and how that translates into an algorithm implementation for $d_{RFL}$.

Regarding the first topic, one should understand that, for \textit{identical} trees $A,B\in\gamma^w_S$, since exists a bijective function $h_V:V_A\longrightarrow V_B$, the matching function $h:E_A\longrightarrow E_B$ can actually be given (informally) by $h(v_1v_2)=h_V(v_1)h_V(v_2)$. This implies that the matching function $h_{(A,B)}$ is \textbf{bijective} and $(E_A=E'_A)\ \wedge\ (E_B=E'_B)$ and, as a consequence (reinforcing that this only holds for $A,B\in\gamma^w_S$ identical) $$d_{RFL}(A,B)= \sum _{e\in E'_A} | w(e) - w(h_{(A,B)}(e))|^k,\ for\ k=1$$
The fact that usually, given the field of application, trees are \textit{roughly identical} influenced that most literature that covers \textit{Robinson-Foulds Length} only considers this part of the distance function to discriminate distance between trees. That can be seen, for example, in \cite{bib:1} which also refers to variations of $d_{RFL}$ by raising every part of the sum by a power of some $k\in\mathbb{N}$, which is the case o Kuhner and Felsenstein (1994).

\begin{figure}[!htb]
  \centering
  \includegraphics[width=0.8\textwidth]{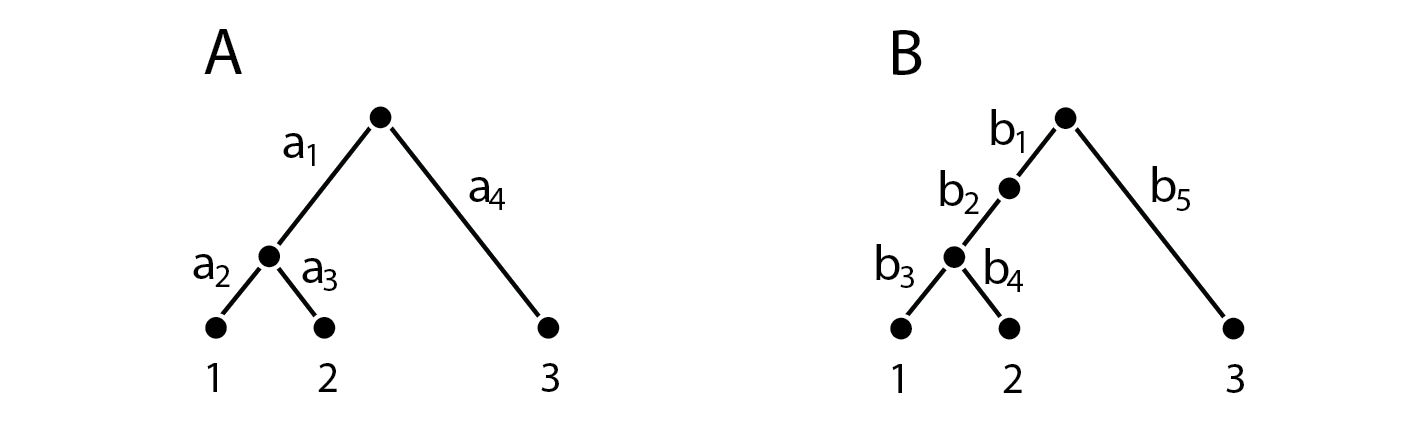}
  \caption[Caption for figure in TOC.]{ Trees $A,B\in\gamma^w_S$ with undefined weights.}
  \label{fig:RFLe}
\end{figure}

One interesting thing to note is the limitations of $d_{RFL}$, since can give multiple results for the same input and it isn't symmetric. Consider the trees $A,B\in\gamma^w_S$ in Figure \ref{fig:RFLe}.

It should be trivial to understand that:
\begin{itemize}
  \item The edge sets $E'_A$ and $E'_B$ equal $E_A$ and $E_B$ respectively;
  \item There are two possible functions $h_{(A,B)}:E_A\longrightarrow E_B$: one that maps $a_1$ into $b_1$ and other that maps $a_1$ into $b_2$;
  \item $h_{(B,A)}:E_B\longrightarrow E_A$ maps $b_1$ and $b_2$ in $a_1$.
\end{itemize}

Given this, one should also realize that we have two different values for $d_{RFL}(A,B)$ depending on the chosen $h_{(A,B)}$ function. These are results of several problems in \cite{bib:12}: Theorem 4 proves the existence of $h$ matching function between identical trees, however, there's no unicity assured; The authors define (differently to Day, Notation \ref{nota:day}) $T'_1$ and $T'_2$ as the trees generated by colapsing edges in $E_1\backslash E'_1$ and $E_2\backslash E'_2$ respectively as identical, but in the example we just exposed that doesn't hold; Definition 5 assumes a unique $h$ between $T'_1$ and $T'_2$ but that might not be the case, as we just exemplified.

But the problems don't end here. We ask the reader to check if the symmetry and identity of indiscernibles holds in $d_{RFL}$ for all $A,B\in\gamma^w_S$, which the second can be easily refuted by considering the weight of every edge in $A$ and $B$ of our example as $1$.

From a computational standpoint, the challenge of implementing $d_{RF}$ and $d_{RFL}$ are aproximately the same, but not without realizing the relation between the $E'_A$ and $E'_B$ edge sets with $C(A,B)$, given $A,B\in\gamma^w_S$. If we take the definition of the \textit{strict consensus tree} (Definition \ref{dfn:scmsct}) for two trees and ask what's its edge set, we understand that it must consist of a set of edges that must be matched in both trees. \textbf{If we assume that there's an one-to-one correspondence} between the \textbf{connected components} (or \textbf{edges}) from the definition of \textit{partitioning function} (Definition \ref{dfn:partfunc}) and the \textbf{clusters} from the \textit{cluster representation} (which may not be the case, as we shown with Figure \ref{fig:RFLe}) we can prove the existence of unique bijective functions between $E'_A$, $E'_B$ and $E_{C(A,B)}$. This implies that the complexity of computing sets $E'_A$ and $E'_B$ is reducted to the complexity of computing the tree $C(A,B)$ which, by William Day's work, we know it's $O(n)$.

\begin{nota}
Let $A\in\gamma^w_S$ and $B$ a subtree of $A$. $S\mid_B$ is the subset of $S$ in which its elements are labels of some vertex in $B$. Also, let $v\in V_A$. We define $A(v)$ as the subtree of $A$ that consists of $v$ and all its lineal descendants.\par
\textbf{Also, for the next proof, for any trees $A,B\in\gamma_S$ we'll denote by $A'$ and $B'$ as the trees generated by collapsing all the edges in $A$ and $B$ that belong in the set $E_A\backslash E'_A$ and $E_B\backslash E'_B$ respectively, and $(A)'$ and $(B)'$ as the cluster representation of $A$ and $B$ respectively.}\par
Assume as well that for every edge $uv\in E_X$ for $X\in\gamma^w_S$, $v$ is deeper than $u$.
\end{nota}

\begin{thm}
Let $A,B\in\gamma^w_S$, $f_A,f_B$ the respective partitioning functions and $C(A,B)$ their strict consensus tree. Assuming $(\dag)$ there's a one-to-one correspondence between edges of $A$ and $B$ and their respective clades and $(\dag\dag)$ for all $a\in E_A$ there's no $a'\in E_A$ such that $S\mid _{A(a)}= S\backslash \big( S\mid _{A(a')}\big)$, there's bijective matching functions $h_{(C(A,B);A')}$ and $h_{(C(A,B);B')}$.
\end{thm}

\begin{proof}

Let $xy\in E_{C(A,B)}$. By definition of the strict consensus tree $C(A,B)$
$$(S\mid_{C(A,B)(y)})\in (A)' \wedge (S\mid_{C(A,B)(y)})\in (B)' $$
by the $(\dag)$ property, we have that exists one and only one $a_1a_2\in E_A$ and one and only one $b_1b_2\in E_B$ such that $$(S\mid_{C(A,B)(y)})=(S\mid_{A(a_2)})=(S\mid_{B(b_2)})$$ which is equivalent to state that $$f_A(a_1a_2)=\{(S\mid_{C(A,B)(y)}),S\backslash (S\mid_{C(A,B)(y)})\}=f_B(b_1b_2).$$
which implies that $a_1a_2$ and $b_1b_2$ are matched edges, hence $a_1a_2\in E'_A$ and $b_1b_2\in E'_B$. Then, we can establish that our matching functions will be such that $h_{(C(A,B);A')}(xy)=a_1a_2$ and $h_{(C(A,B);B')}(xy)=b_1b_2$.

We'll now prove that $h_{(A;C(A,B))}$ is bijective (the proof for $h_{(C(A,B);B)}$ will be left for the reader).
Let $a_1a_2=a$ and $a'_1a'_2=a'$ be edges of $A$. If $h_{(A;C(A,B))}(a)=h_{(A;C(A,B))}(a')$ then $f_A(a)=f_A(a')$. Since there's a one-to-one correspondence between the edges of $A$ and it's clades $(\dag)$ and it can't be the case that $S\backslash (S\mid_{A(a_2)})= S\mid_{A(a_2')}$ $(\dag\dag)$ we have that $a$ must be equal to $a'$, proving the injectivity of $h_{(A;C(A,B))}$. For surjectivity, let $xy\in E_{C(A,B)}$. By definition of strict consensus tree, we have that $S\mid _{C(A,B)(y)}\in (A)'$. We have that, by $(\dag)$, $\exists a_1a_2\in E_A$ such that $(S\mid_{A(a_2)})=(S\mid_{C(A,B)(y)})$. This means that $f_A(a_1a_2)=f_{C(A,B)}(xy)$ hence $h_{(A;C(A,B))}(a_1a_2)=xy$, proving surjectivity.
\end{proof}

Regarding the \textbf{discrimanory power}, one should understand that, other than the characteristics inerited from the fact that $d_{RF}$ is purely a topological measure and $d_{RFL}$ considers branch length, \textbf{Robinson Foulds} inspired metrics (at least the ones discussed here) are really sensible to the scalability of $S$ \cite{bib:4}.

\begin{figure}[!htb]
  \centering
  \includegraphics[width=0.8\textwidth]{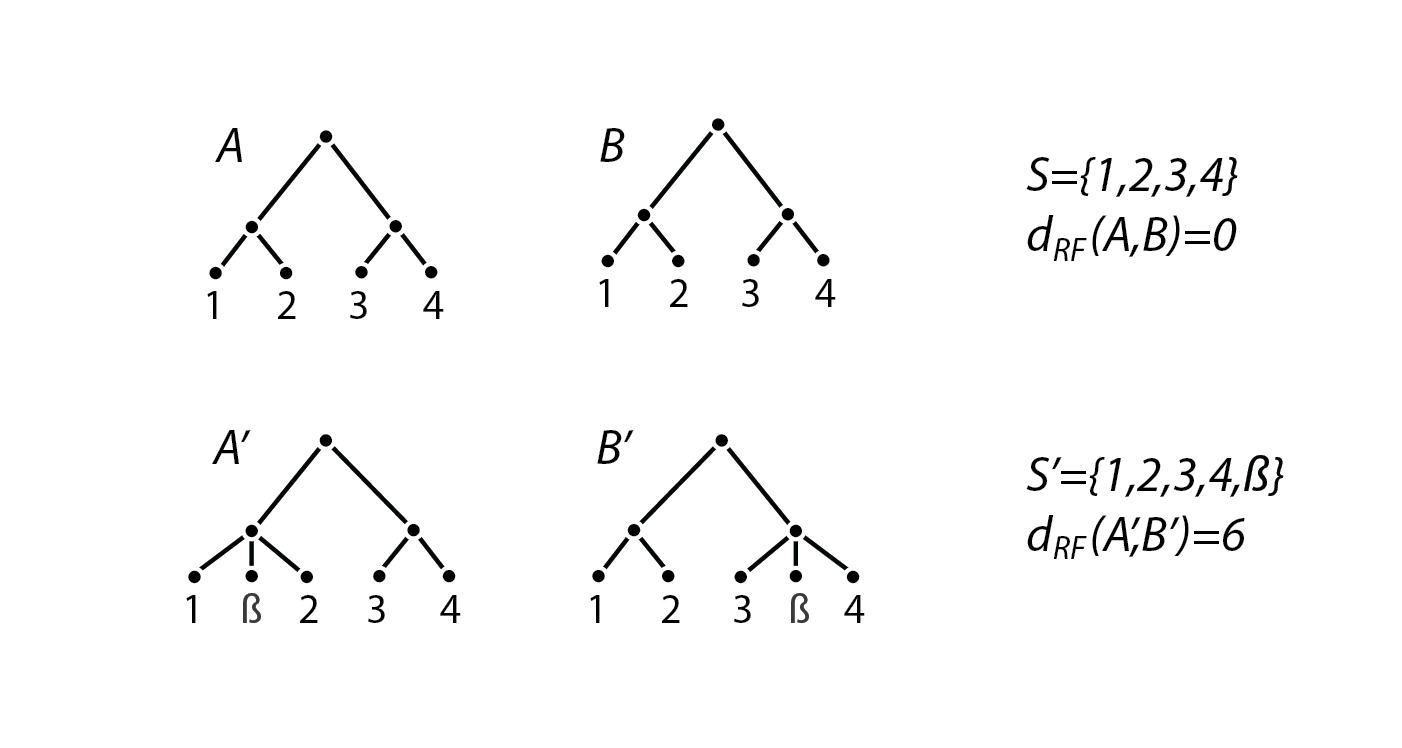}
  \caption[Caption for figure in TOC.]{ Trees in $\gamma_S$ and $\gamma_{S'}$ and the \textit{RF} distance between them. }
  \label{fig:RFOutl}
\end{figure}

For instance, if we have $A,B\in\gamma_S$ and $d_{RF}(A,B)=k$ for some $k$, having $A',B'\in\gamma_{S'}$ such that $A$ and $B$ are \textit{non-trivial} subtrees of $A'$ and $B'$ respectively (meaning: $S\varsubsetneq S'$, $V_A\varsubsetneq V_{A'}$, $V_B\varsubsetneq V_{B'}$, $E_A\varsubsetneq E_{A'}$, $E_B\varsubsetneq E_{B'}$), there's no direct relation between $d_{RF}(A',B')=k$ whatsoever, since all clades that were shared between $A$ and $B$ can now be different. The same applies for $d_{RFL}$. To illustrate this, consider the Figure \ref{fig:RFOutl} for $d_{RF}$.

So, would be reasonable to assume that if one wants to adopt an iterative method for his problem on size of $|S|$, \textit{RF} based metrics wouldn't be a good approach.

Not only the scalability of $S$ is a problem, but moving a single leaf could lead to great discrepancies in the distance value. All around, \textit{RF} metrics end up beeing a fairly unstable metric to work with, althrough serving its purpose for distinguishing trees \cite{bib:1}. All due to the fact that it's a metric that works with shared clades: if a leaf is replaced on the tree, all the clades which it belongs to will necessarily be different. Another consequence of this fact is that \textit{RF} will overperform in close to resolved trees other than to unresolved ones \cite{bib:1,bib:16} (beeing a unresolved tree a tree which the internal nodes have mostly degree greater than $3$).

%%%%%%%%%%%%%%%%%%%%%%%%%%%%%%%%%%%%%%%%%
%%%%%%%%%%%Trip, TripL, Quart%%%%%%%%%%%%
%%%%%%%%%%%%%%%%%%%%%%%%%%%%%%%%%%%%%%%%%

\subsection{Quartets, Triplets and Triplets Length }
\label{subsection:trip}

The methods that we're about to introduce are recent compared to previous ones and results on complexity differ according to the features of the considered data structures. First approach was made in 1985 by George F. Estabrook, F. R. McMorris and Christopher A. Meacham with the publication of the \textit{Quartets distance} in \cite{bib:16} and wasn't until 11 years later when Douglas E. Critchlow, Dennis K. Pearl and Chunlin Qian provided a formal definition for the \textit{Triplets distance} in \cite{bib:14} which is heavily inspired by the former. In 2014 Mary K. Kuhner and Jon Yamato made a study to compare practical perfomance of a variety of different metrics \cite{bib:1} and for that matter thought it was interesting to consider a metric that would take the topology analysis properties of the latter but consider branch length as well.

\begin{figure}[!htb]
  \centering
  \includegraphics[width=0.8\textwidth]{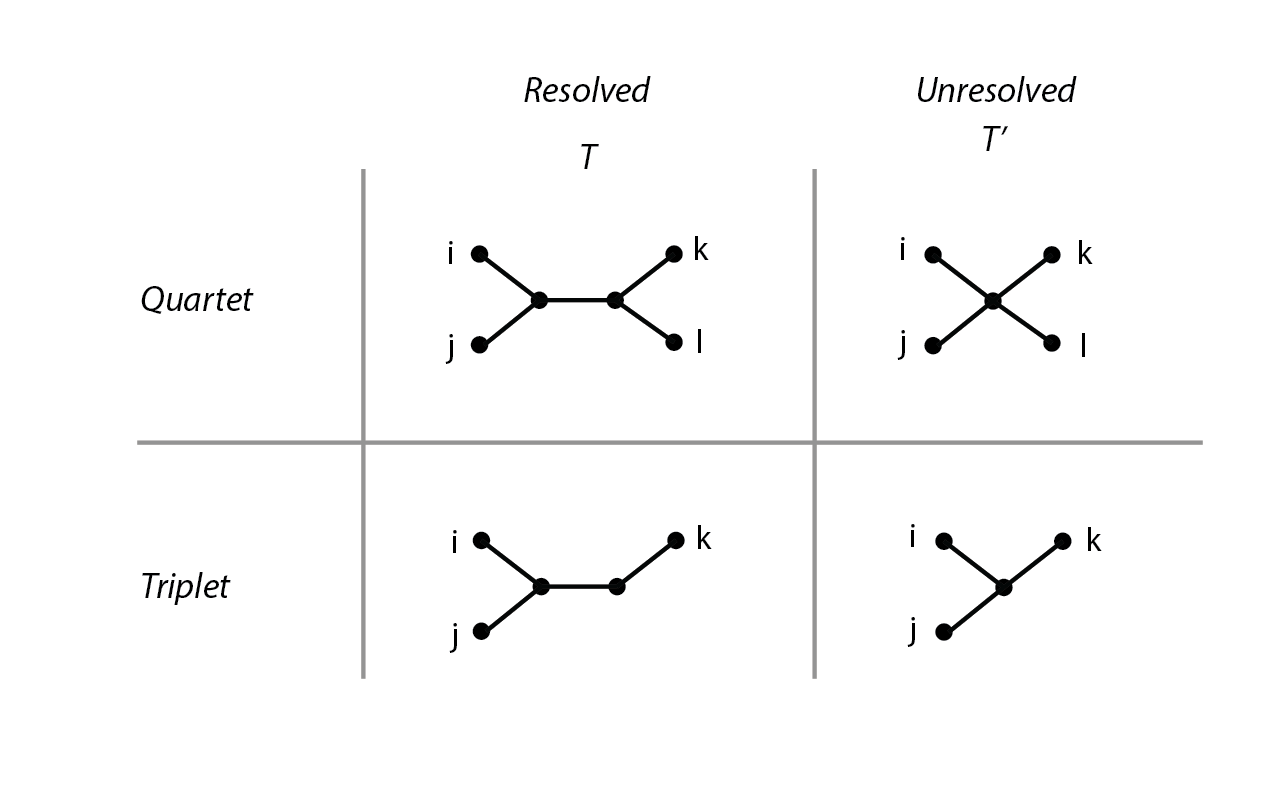}
  \caption[Caption for figure in TOC.]{Resolved and unresolved quartets and triplets.}
  \label{fig:QTTopo}
\end{figure}

The inicial thought behind Estabrook, \textit{et al.} \textit{Quartet Distance} was how phylogenetic tree aggreement behaves with respect to the topologic aspect of the branching alone, disregarding direction. However, and as stated previously, the problem differs regarding the structure we're applying the distance: \textit{binary} trees only lead to \textbf{\textit{resolved quartets/triplets}} while \textit{non-binary} can lead also to \textbf{\textit{unresolved quartets/triplets}}. These quartets/triplets can be consulted in Figure \ref{fig:QTTopo}.

\begin{dfn}\label{dfn:quart1}
{\textbf{Quartet Distance} (informal)}\\
Let $A,B\in \gamma_S$, $V_X$ the vertex set for any tree $X\in \gamma_S$, $T$ the quartet depicted in Figure \ref{fig:QTTopo} for the resolved case and $\bar{d}(a,b)$ the usual edge distance between vertices $a,b\in V_X$ in the tree $X$. The \textbf{Quartet distance} $d_Q$ consists in:
\begin{itemize}
 \item Consider every subset $S'$ of size $4$ from the set of leaves $S$;
 \item Build maps $\sigma_A$ and $\sigma_B$ (in case they exist) from the vertex set of the subtrees of $A$ and $B$ generated by considering only edges connecting leaves from $S'$ (that we will designate as $A\mid _{S'}$ and $B\mid _{S'}$) to the vertex set of $T$ such that, given $X\in\{A,B\}$, for every $v_1,v_2,v_3\in V_{X\mid_{S'}}$: $$\bar{d}(v_1,v_2)\leq \bar{d}(v_1,v_3) \Rightarrow \bar{d}(\sigma_X(v_1),\sigma_X(v_2)) \leq \bar{d}(\sigma_X(v_1),\sigma_X(v_3))$$
 
 \item Build partitions $T^p_{A\mid_{S'}}$ and $T^p_{B\mid_{S'}}$ for the labels $S'$ such that, for all $X\in\{A, B\}$, $v_1,v_2\in V_{X\mid_{S'}}:\ v_1, v_2\ \mathrm{labeled\ vertices}$: $$(\bar{d}(\sigma_X(v_1),\sigma_X(v_2))=2)\Rightarrow (\{label(v_1),label(v_2)\}\in T^p_{X\mid_{S'}})$$
 \item If $T^p_{A\mid_{S'}}\neq T^p_{B\mid_{S'}}$ or exactly one of the maps $\sigma_A$ and $\sigma_B$ doesnt exist, account $1$ for the quartet distance $d_Q$. 
\end{itemize}
\end{dfn}

If exactly one of the mappings $\sigma_A$ or $\sigma_B$ doesn't exists, it means one of the quartets is unresolved in tree $A$ or $B$, so they necessarily differ. If both mappings $\sigma_A$ and $\sigma_B$ don't exist, it means that in both trees, $A$ and $B$, the quartet is unresolved, hence, they agree.

However, in article \cite{bib:11} for \textit{Triplet distance} is presented an informal definition that we find more suitable to understand the concept behind these two metrics (\textit{Triplets} and \textit{Quartet distances}), however, this falls short by semantic reasons.

\begin{dfn}{\textbf{Quartet and Triplet distance} (informal)}\\
Let $A,B\in\gamma_S$. Consider $S'$ as every subset of $S$ of size $k$ and the indicator function defined as
$$I_{S'}=
\begin{cases} 
      1 & \mathrm{if\ labels\ on\ }S'\mathrm{\ have\ different\ }\mathrm{subtrees\ in\ }A\mathrm{\ and\ }B \\
      0 & \mathrm{otherwise} 
\end{cases}$$

Then, the \textbf{Triplet distance} $d_{Trip}(A,B)$ (for $k=3$) and \textbf{Quartet distance} $d_{Q}(A,B)$ (for $k=4$) are given by $$\sum_{S'\subset S\ :\ |S'|=k} I_{S'}$$

\end{dfn}

The problem with this definition is that the \textit{different subtrees} referred in indicator $I_{S'}$ isn't the straightforward notion of \textit{different}. Actually, to achieve the comparison between subtrees that Critchlow, \textit{et al.} (1996) (from \cite{bib:14}) are referring, one would need to erase every label information on $A$ and $B$ other than $S'$ and then consider the trees $A'$ and $B'$ generated with the smallest amount of contractions of Bourque $\alpha$ from $A$ and $B$ with same topology as $T$ or $T'$ (depending on which one requires least contractions) from Figure \ref{fig:QTTopo} and at most $1$ label for each leaf, $0$ labels for internal nodes. \textbf{To obtain the \textit{Triplet distance} from the definition \ref{dfn:quart1} (of the \textit{Quartet distance}) we need to consider subsets $S'$ of size $3$ instead of size $4$ and consider $T$ from Figure \ref{fig:QTTopo} for triplets instead of quartets}.

In the article of Kuhner \textit{et al.} \cite{bib:1} a new metric was considered to maintain $d_{Trip}$'s topologic discriminatory power and also acount for weight information. The main idea was to account the path weight between the tips with labels from $S'$, but do it when the topology of the \textit{``subtrees"} is \textbf{equal} rather than different. Considering the definition from Critchlow, \textit{et al.} (1996) \cite{bib:14} as a starting point, we define the \textit{Triplet Length distance} bellow:

\begin{dfn}{\textbf{Triplet Length distance} (informal)}\\
Let $A,B\in\gamma_S$. Consider $S'=\{i,j,k\}$ as every subset of $S$ of size $3$, $v_{(i,X)}\in V_X$ the singular vertex in $X$ with label $i$ and the indicator functions $I_{S',1}$ and $I_{S',2}$ defined as
$$I_{S',1}=
\begin{cases} 
      0 & \mathrm{if\ labels\ on\ }S'\mathrm{\ have\ different}\\ & \mathrm{subtrees\ in\ }A\mathrm{\ and\ }B \\
      |\bar{d_l}(v_{(i,A)},v_{(j,A)})-\bar{d_l}(v_{(i,B)},v_{(j,B)})| & \mathrm{otherwise} 
\end{cases}$$

$$I_{S',2}=
\begin{cases} 
      0 & \mathrm{if\ labels\ on\ }S'\mathrm{\ have\ different}\\ & \mathrm{subtrees\ in\ }A\mathrm{\ and\ }B \\
      |\bar{d_l}(v_{(i,A)},v_{(k,A)})-\bar{d_l}(v_{(i,B)},v_{(k,B)})| & \mathrm{otherwise} 
\end{cases}$$

Where $\bar{d_l}(v,u)$ is the weight of the path between $u$ and $v$. Then, the \textbf{Triplet Length distance} is defined as $$d_{TripL}(A,B)=\sum_{S'\subset S\ :\ |S'|=3} (I_{S',1}+I_{S',2})$$

\end{dfn}

This metric is fairly recent, considered for the purposes of the study in \cite{bib:1} and its relevance is underwhelming (as we will state and can be checked on the results of the article).

When comes to \textbf{implementation}, an easy way to structure all the possible situations when analysing the \textit{quartets} or \textit{triplets} is depicted in table \ref{tab:QTpos}.

\begin{table}[]
\centering

\label{tab:QTpos}
\begin{tabular}{l|l|l}
                          & Resolved      & Unresolved           \\ \hline
\multirow{2}{*}{Resolved} & $A$: Agree    & \multirow{2}{*}{$C$} \\ 
                          & $B$: Disagree &                      \\ \hline
Unresolved                & $D$           & $E$                 
\end{tabular}
\caption{Categorization of the different types of Quartets/Triplets, necessary for the computation of the metrics.}

\end{table}

And actually, as Brodal, \textit{et al.} specifies in \cite{bib:13}, a paper focused on efficient algorithms to compute Triplet and Quartet distance, the lines and rows of this table can be calculed in $O(n)$ time through \textit{dynamic programming}. Since the quartet and triplet distance consists only in adding up $B+C+D$, then the main idea is to find a way of computing $E$ and $A$, and that's the focus of \cite{bib:13}.

The conclusion is that the algorithm for finding $A$ and $E$ differs in complexity depending on the structure: For rooted trees and triplets, $A$ and $E$ can be computed in time $O(nlog(n))$; unrooted trees and quartets, $A$ can be computed in $O(nlog(n))$ and $B$ in $O(dnlog(n))$ where $d$ is the maximum vertex degree of any node in the two trees.

As for the \textbf{discriminatory power} is interesting to understand how these metrics relate to the scalability of $S$, where \textit{RF} metrics underperforms. Actually, if we consider $A_0,B_0\in\gamma_{S_0}$ and a chain of non-trivial \textit{supertrees} $A_i,B_i\in\gamma_{S_i}$ for $i\in\mathbb{N}$ where $S_i\varsubsetneq S_{i+1}$ and $A_i$ and $B_i$ are non-trivial subtrees of $A_{i+1}$ and $B_{i+1}$ respectively, it's reasonable to understand that $d_{Q}(A_i,B_i)\leq d_{Q}(A_{i+1}, B_{i+1})$ and $d_{Trip}(A_i,B_i)\leq d_{Trip}(A_{i+1}, B_{i+1})$. This is due to the fact that these metrics value relations between subsets of size $k$ ($k=4\vee k=3$), and any change done to leaves will only affect the part of the sum related to quartets/triplets where that leaf is contained.

However, regarding its practical performance on the main field of application (phylogenetics) \cite{bib:1}, \textit{Quartet} based metrics didn't performed well, and according to Kuhner, \textit{et al.} is due to these distances beeing more sensible to the bottommost branchings of the tree, once a large portion of these branches are contained in these branchings. This last conclusion might be too specific for the example but, if that's the case, will lead us to believe that \textit{Quartet} based metrics will overperform in unresolved trees over resolved ones (beeing a unresolved tree a tree which the internal nodes have mostly degree greater than $3$).

\subsection{Geodesic distance}
\label{subsection:geo}

The most recent metric that brought original insight for the classification problem was product of Louis J. Billera, Susan P. Holmes and Karen Vogtmann. Since the classical problem of phylogeny is to find a tree which is more consistent with the \textit{taxonomical} data, knowing how much the calculated tree is correct becomes also a statistical problem: would a small change in the data will result in a change of choice in the resulting tree (as we saw this is a limitation of \textit{Robinson-Foulds metric}). The fact that this can be considered as a problem in the estimation process lead various authors to suggest to partition the space of trees into regions, and that's what Billera, \textit{et al.} specifies in \cite{bib:15}. The \textbf{Geodesic distance} is a distance built over the space of trees.

That leaves us with the task of primarely formalizing the space where the distance will be built. Take in consideration that an \textbf{internal edge} is any edge that is not connected to a leaf of a tree.

\begin{dfn}  \textbf{Space of trees with $n$ labels, $\mathcal{T}_n$}\\
Consider $S$ a set of labels and $|S|=n$. For every non-identical binary tree $B_i\in\gamma_S$ (there is a total of $(2n-3)!!$ non-identical binary trees, where $!!$ stands for the \textbf{double factorial}) generate an $(n-2)$-dimensional space $\mathcal{T}^o_B$ (that we designate as \textbf{orthant}) such that every component $c_e$ is identified to one (and only one) internal edge $e\in E_B$ and takes values between $[0 , \infty [$. For all pairs of spaces $\mathcal{T}^o_{B_1}$ and $\mathcal{T}^o_{B_2}$ (with $e_1\in E_{B_1}$ and $e_2\in E_{B_2}$) identify components $c_{e_1}$ and $c_{e_2}$ if and only if the cluster representation of the clades associated with the removal of $e_1$ and $e_2$ from their respective trees match, that is, if $(C_{e_1})'=(C_{e_2})'$ (or, according to Definition \ref{dfn:medg}, $e_1$ and $e_2$ are matched edges). The \textbf{Space of trees with $n$ labels} is the result all the orthants $\mathcal{T}^o_B$ with this identification.
A point $(t_1,t_2,...,t_k)\in\mathcal{T}_n$ specifies a unique tree $A\in \gamma^w_S$ with internal edges $e_i$ such that $w(e_i)=t_i$ for all $t_i\neq 0$. 
\end{dfn}

For better understanding, consider the following two examples:

\begin{ex} \textbf{Space of trees with $3$ labels}, $\mathcal{T}_3$ \\
  The topology of binary trees with $3$ labels is unique, so, if we consider the set $S=\{1,2,3\}$, there are only $3$ non-identical binary trees ($(2\times 3-3)!!=3$), depicted Figure \ref{fig:T3rep}. Each of those will generate an $1$-dimensional space, that will meet by their origin.

\begin{figure}[!htb]
  \centering
  \includegraphics[width=0.8\textwidth]{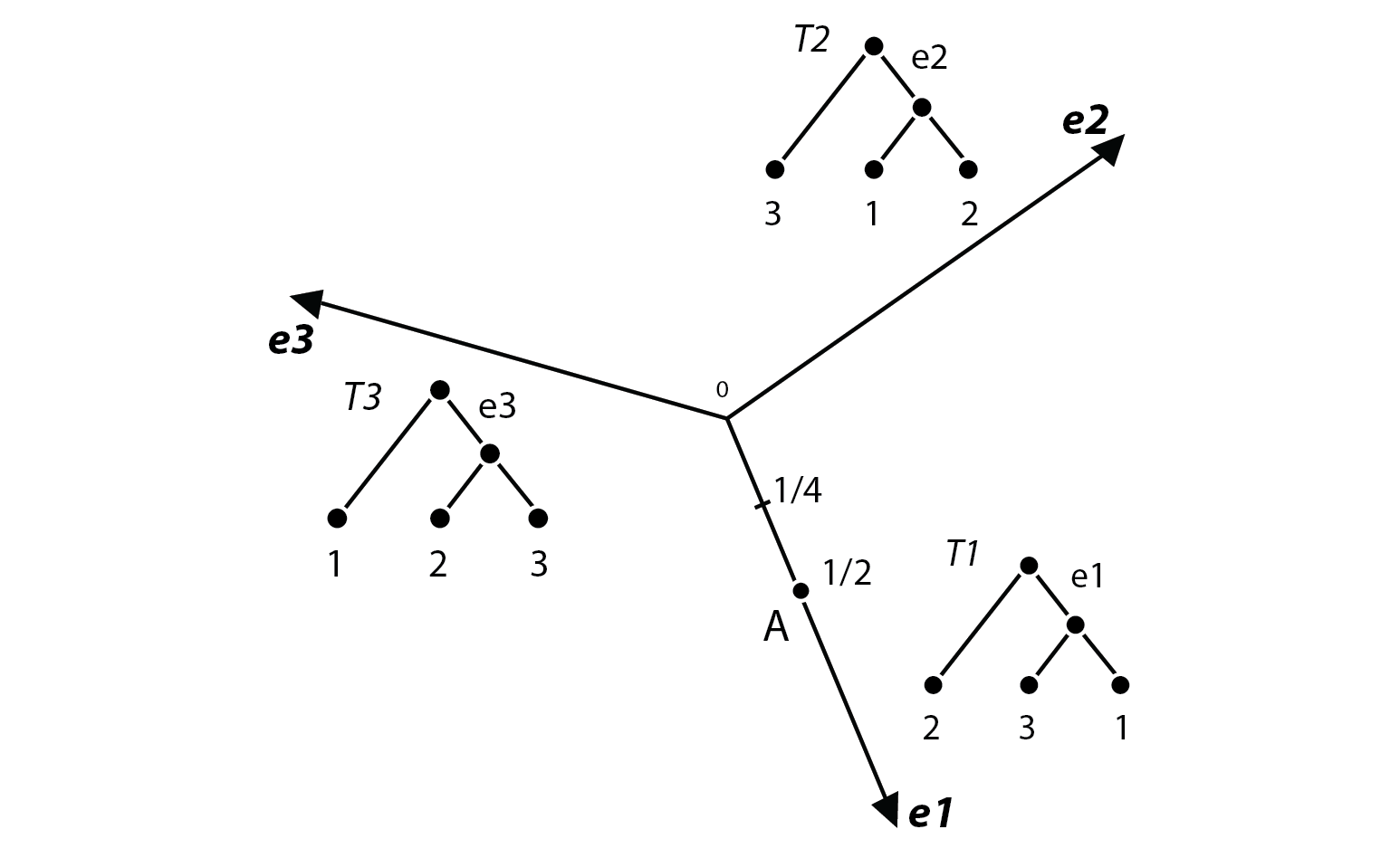}
  \caption[Caption for figure in TOC.]{$\mathcal{T}_3$}
  \label{fig:T3rep}
\end{figure}

The tree at point $A$ is a tree whose topology and labeling are equal to those of $T1$, however, $w(e1)=1/2$.

\end{ex}

\begin{ex} \textbf{Space of trees with $4$ labels}, $\mathcal{T}_4$ \\
  Let $S=\{ 1,2,3,4\}$. The dimension of each orthant will be $(n-2)=(4-2)=2$, meaning that each binary tree will have exactly two internal edges. Also, there are $(2\times 4 - 3)!!= 5!! = 5\times 3\times 1 = 15$ different binary trees. Take the next figure as an example of one of its composing orthants $\mathcal{T}^o_B$, for the depicted $B\in\gamma_S$.

\begin{figure}[!htb]
  \centering
  \includegraphics[width=0.8\textwidth]{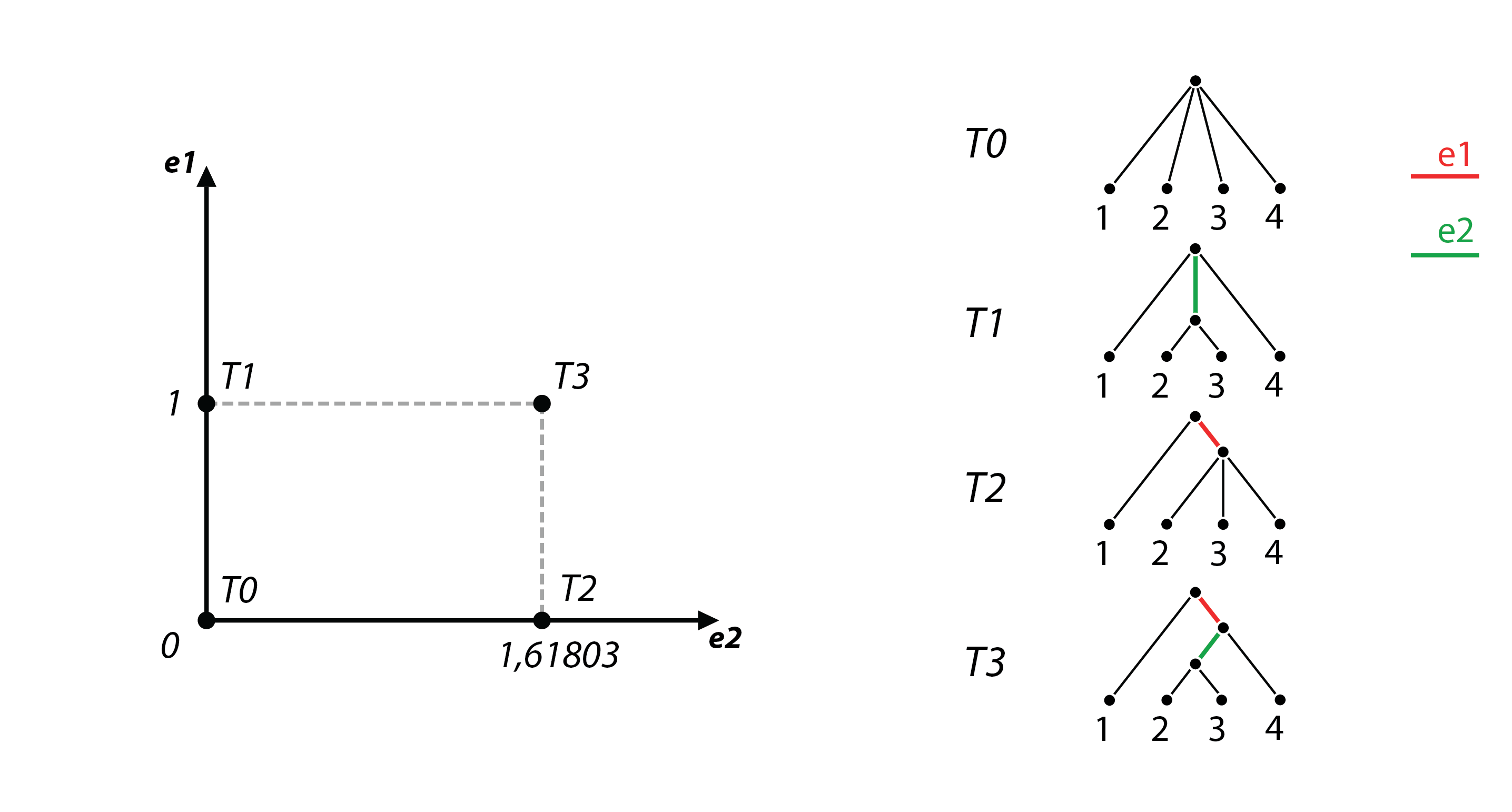}
  \caption[Caption for figure in TOC.]{$\mathcal{T}^o_B$, where $B\in\gamma_S$ is the binary tree identical to $T3$. Given $e_1,e_2\in E_{T3}$, we have that $w(e1)=1$ and $w(e2)=1.61803$.}
  \label{fig:T4Orep}
\end{figure}

As a matter of fact, for the topology of trees with $4$ labels we have five different candidates that can be seen on the Figure \ref{fig:T4O5rep}. And also, the identification of edges will be such that the five orthants associated with these trees will be joined by components two by two.

\begin{figure}[!htb]
  \centering
  \includegraphics[width=0.9\textwidth]{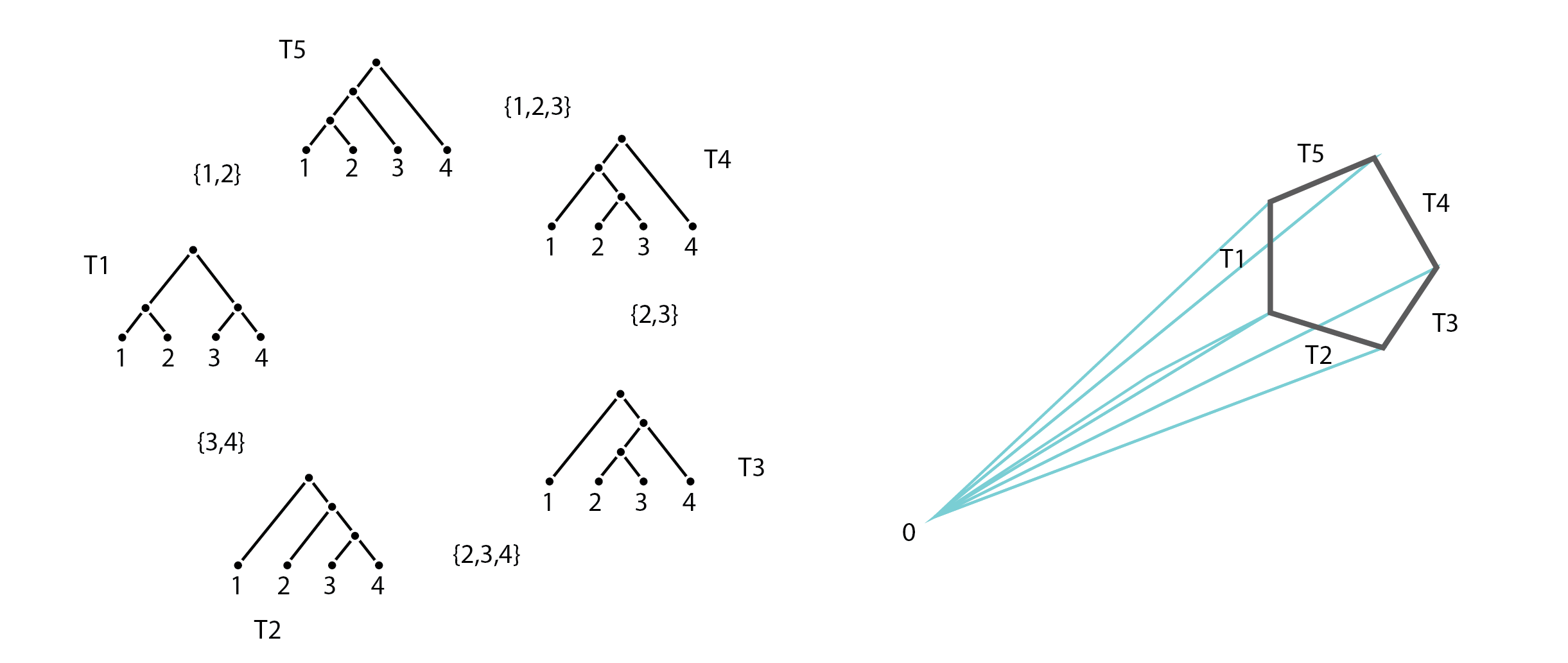}
  \caption[Caption for figure in TOC.]{The five different topologies for trees with $4$ labels and the representation of respective orthants after component identification. It's important to note that, since $\mathcal{T}_4$ isn't presentable in $\mathbb{R}^3$ we will loosen our representations a bit for better understanding. Also note that each triangle in this figure it's actually just a region of each orthant, equivalent to the one in Figure \ref{fig:T4Orep}: the five orthants are actually an infinite \textit{`cone"} on the one depicted (with cone point the origin). }
  \label{fig:T4O5rep}
\end{figure}

Since we can permutate the labels on trees, $\mathcal{T}_4$ will actually be composed of several of these spaces, until we cover all $15$ binary trees. If we consider all $15$ orthants and do the respective identifications we will end up with what can be seen in Figure \ref{fig:T4rep}.

\begin{figure}[!htb]
  \centering
  \includegraphics[width=0.8\textwidth]{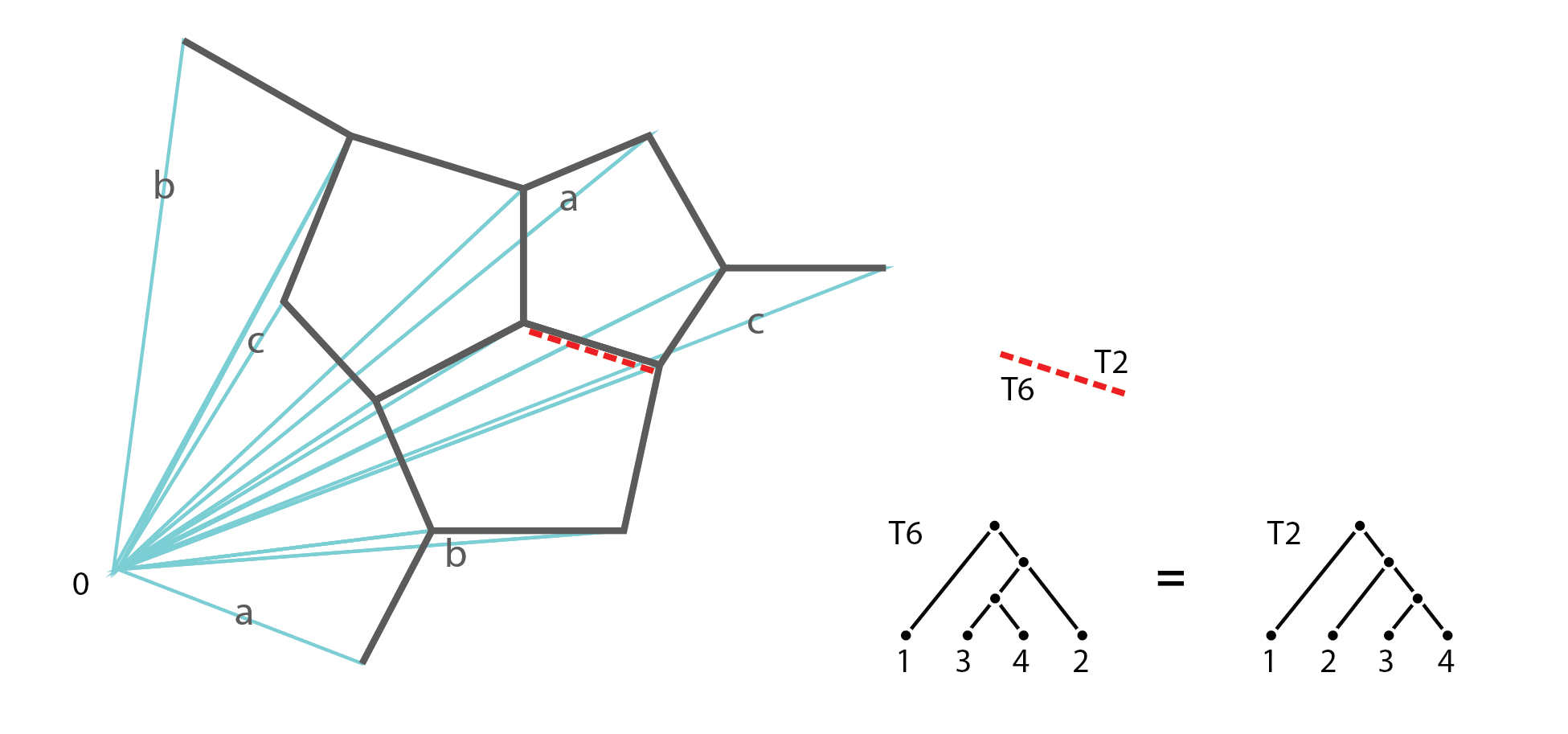}
  \caption[Caption for figure in TOC.]{ Representation of $\mathcal{T}_4$, given the labeled edges are identified as the same components. The trees $T2$ and $T6$, althrough belong to different pentagons (that are associated with different label permutations), are actually the same tree.}
  \label{fig:T4rep}
\end{figure}

It's important to understand that, even though the other ``infinite hexagonal cones" are copies of the first considered in Figure \ref{fig:T4O5rep}, when we match their components, trees with suposely different topologies are actually the same with coinciding orthants, given the label permutation considered. That would lead us to believe that we could identify one $5$-sided polygon for each permutation of the set $S$, and since $|S|=4$ we would have $4!=24$ polygons, instead of $12$ that are identified on the Figure \ref{fig:T4Cycles}.

\begin{figure}[!htb]
  \centering
  \includegraphics[width=1\textwidth]{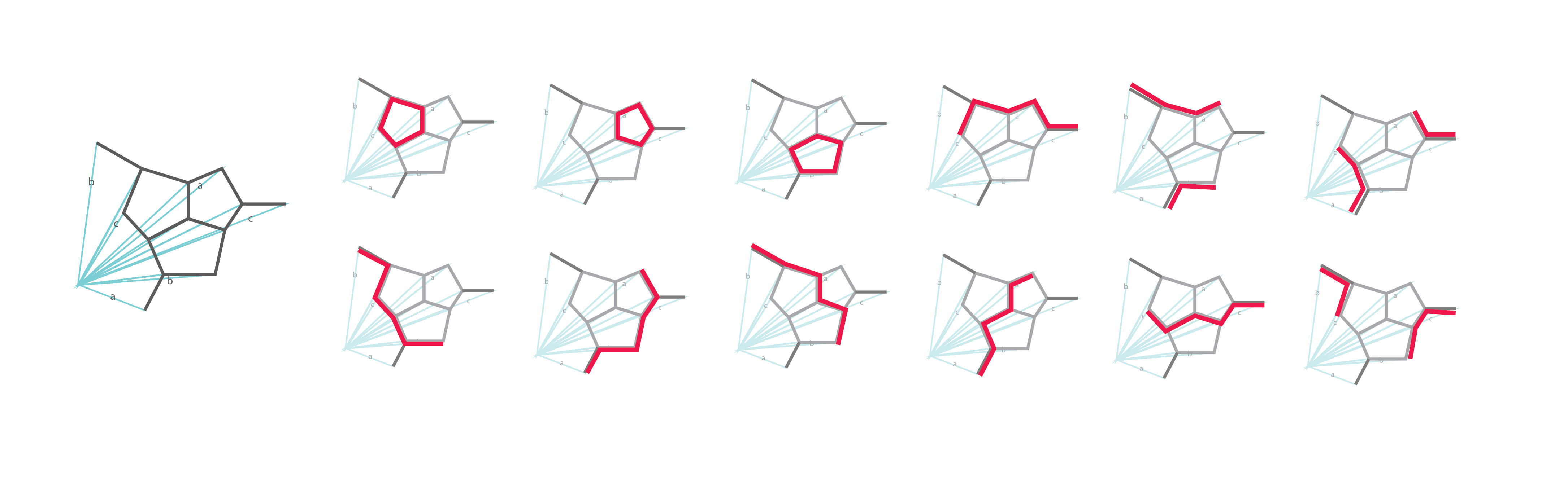}
  \caption[Caption for figure in TOC.]{ All the $5$-polygons in $\mathcal{T}_4$. These actually match the $5$-cycles in \textit{Petersen Graph}.}
  \label{fig:T4Cycles}
\end{figure}

However, if we take a good look at Figure \ref{fig:T4O5rep} we notice that the permutation of labels $(2\leftrightarrow 3;\ 1\leftrightarrow 4)$ would lead to the same set of $5$ orthants: not only those trees are identical, but the same tree. So each $5$-side polygon represents, actually, two permutations of labels instead of one.

\end{ex}

Now that we defined this space of trees, we are left to understand what's the metric associated with this space. Its construction leads us to conclude that it already comes equipped with a natural distance function: as a matter of fact, this space is made up of \textit{standard Euclidean} orthants. So, the distance between two points (or trees) in the same orthant will be the usual Euclidean distance. If two points are in different orthants, we can build a path between them that is a sequence of straight segments, each one laying in a single orthant. We can then measure the length of the path by adding up the lengths of the segments. Lets denote this distance in $\mathcal{T}_{|S|}$ space as $d_{\mathcal{T}_{|S|}}(A,B)$.

The existence of this path along orthants is given by the fact that, for all $n$, $\mathcal{T}_n$ is a space with \textit{non-positive curvature} (proof of Lemma 4.1 in \cite{bib:15}) and follows from Gromov, 1987 \cite{bib:27} that all these spaces have an unique shortest path connecting any two points called \textbf{geodesic}, hence the name of the metric.

\begin{dfn} \textbf{Geodesic distance}\\
Let $A,B\in\gamma^w_S$, $\mathcal{T}_{|S|}$ the space of trees with $|S|$ labels and $d_{\mathcal{T}_{|S|}}$ the associated distance function. Then, the \textbf{Geodesic Distance} $d_{Geo}(A,B)=d_{\mathcal{T}_{|S|}}(A,B)$.
\end{dfn}

Althrough Billera, \textit{et al.} \cite{bib:15} aproaches how to calculate this metric (and we recommend the article for more insight), it's far from describing an implementation. Far from previous approaches that lead to exponential time algorithms, Megan Owen and J. Scott Provan (2009) described a polynomial time algorithm to compute this distance in \cite{bib:9}, which we will approach next.

\begin{nota}
Let $T=(V,E,S,w)\in\gamma^w_S$. Each internal edge $e\in E$ defines a partition of $S$ into two sets, determined by the cluster representation of the two connected components from the removal of the edge $e$ from $T$. This partition, in \cite{bib:9}, goes by the name as \textbf{split} and is denoted as $X_e|\bar{X_e}$, which $X_e$ and $\bar{X_e}$ beeing its sets.
\end{nota}

\begin{dfn} \textbf{(Compatible Splits)}
Let $T \in \gamma^w_S$  and $e,f \in E_T$. We say that splits $X_e|\bar{X_e}$ and $X_f|\bar{X_f}$ are \textbf{compatible} if at least one of the sets $X_e\cap X_f$, $X_e\cap \bar{X_f}$, $\bar{X_e}\cap X_f$, $\bar{X_e}\cap\bar{X_f}$ is empty.
\end{dfn}

\begin{dfn} \textbf{(Compatible Edge Sets)}\\
Let $T_1,T_2 \in \gamma^w_S$ and $A \subseteq E_{T_1}$ and $B \subseteq E_{T_2}$. 
\begin{itemize}
  \item A single set $A$ is a \textbf{compatible set} if, for all $a_1,a_2\in A$, the splits $X_{a_1}|\bar{X_{a_1}}$ and $X_{a_2}|\bar{X_{a_2}}$ are compatible;
  \item We say that $A$ and $B$ are \textbf{compatible sets} if for every $a\in A$ and $b\in B$ the splits $X_a|\bar{X_a}$ and $X_b|\bar{X_b}$ are \textbf{compatible}.
\end{itemize}
\end{dfn}

The following theorem is important for the existence of the space of trees as it is, as a matter of fact is what stops it from colapsing. It can be found on \textit{Phylogenetics} by Charles Semple and Mike Steel \cite{bib:17} and its proof takes a whole section, but the statement was originally work of Peter Buneman in 1971.

\begin{thm} (\textbf{Split Equivalence Theorem}, \cite[Theorem 3.1.4]{bib:17})\\
Let $\mathcal{C}$ be a set of splits. $\mathcal{C}$ uniquely defines a tree if and only if $\mathcal{C}$ is a \textbf{compatible set}.
\end{thm}

\begin{dfn} \textbf{(Path Space and Path Space Geodesic)}\\
Let $T_A,T_B\in\gamma^w_S: T_A, T_B\ \mathrm{non-identical}$ and $\mathcal{A}=(A_1,...,A_k)$ and $\mathcal{B}=(B_1,...,B_k)$ partitions of $E_{T_A}$ and $E_{T_B}$ such that the pair $(\mathcal{A},\mathcal{B})$ satisfies:
\begin{itemize}
  \item[] \textbf{(P1)} For each $i>j$, $A_i$ and $B_j$ are compatible sets.
\end{itemize}
Then, for all $1\leq i \leq k$, $B_1\cup ...\cup B_i\cup A_{i+1}\cup ... \cup A_k$ is a compatible set, hence, from the \textbf{splits equivalence theorem}, uniquely defines a binary $|S|$-tree $T_i$ and an associated orthant $\mathcal{T}^o_{T_i}$. The connected space $\mathcal{P}=\bigcup^k_{i=1} \mathcal{T}^o_{T_i}$ is the \textbf{path space} with support $(\mathcal{A},\mathcal{B})$ and the shortest path from $T_A$ to $T_B$ cointained in $\mathcal{P}$ the \textbf{path space geodesic} for $\mathcal{P}$.
\end{dfn}

\begin{thm} (From Billera \textit{et al.} \cite[Proposition 4.1]{bib:15})\\
For trees $T_1,T_2\in\gamma^w_S$ with disjoint edge sets, the geodesic between $T_1$ and $T_2$ is a path space geodesic for some path space between $T_1$ and $T_2$.
\end{thm}

For the a set of edges $A$ we use the notation $||A||=\sqrt{\sum_{e\in A} w(e)^2}$ to denote the norm of the vector whose components are lengths (weights) of the edges in $A$.

\begin{dfn} \textbf{(Proper Path Space and Proper Path)}\\
Let $T_1,T_2\in\gamma^w_S$ and $\Gamma$ the geodesic in $\mathcal{T}_{|S|}$ between $T_1$ and $T_2$. Then, $\Gamma$ can be represented as a path space geodesic with support $\mathcal{A}=(A_1,...,A_k)$ of $E_{T_1}$ and $\mathcal{B}=(B_1,...,B_k)$ of $E_{T_2}$ which satisfy \textbf{P1} and the following additional property:
\begin{itemize}
  \item[] \textbf{(P2)} $\frac{||A_1||}{||B_1||} \leq \frac{||A_2||}{||B_2||} \leq ... \leq \frac{||A_k||}{||B_k||}$
\end{itemize}
We call a path space satisfying conditions \textbf{P1} and \textbf{P2} a \textbf{proper path space} and the associated path space geodesic a \textbf{proper path}.
\end{dfn}

\begin{thm} (From Owen \textit{et al.} \cite[Theorem 2.5]{bib:9})\\
Given $T_1,T_2\in\gamma^w_S$ A \textbf{proper path} $\Gamma$ between $T_1$ and $T_2$ with support $(\mathcal{A},\mathcal{B})$ is a geodesic if and only if $(\mathcal{A},\mathcal{B})$ satisfy the property:
\begin{itemize}
  \item[] \textbf{(P3)} For each support pair $(A_i,B_i)$ there is no non-trivial partitions $C_1\cup C_2$ for $A_i$ and $D_1\cup D_2$ for $B_i$ such that $C_2$ is compatible with $D_1$ and $\frac{||C_1||}{||D_1||}<\frac{||C_2||}{||D_2||}.$
\end{itemize}
\end{thm}

One intuitive path between any two trees that will be usefull for the algorithm implementation as the starting point is the \textbf{\textit{cone path}}. This is path that connects the two trees through two straight line segments through the origin of our space $\mathcal{T}_n$. The \textit{cone path} will function as our starting point with support $(\mathcal{A}^0,\mathcal{B}^0)$ that vacuously satisfies (P1) and (P2). The algorithm goes as follows:

\begin{alg}\textbf{(Geodesic Algorithm)}
\begin{itemize}
  \item[]\textbf{Input}: $T_1,T_2\in\gamma^w_S$;
  \item[]\textbf{Output}: The path space geodesic between $T_1$ and $T_2$.
  \item[]\textbf{Initialize}: $\Gamma^0$ = cone path between $T_1$ and $T_2$ and support $(\mathcal{A}^0,\mathcal{B}^0)=((E_{T_1}),(E_{T_2}))$.
  \item[]\textbf{Step}: At stage $l$, we have proper path $\Gamma^l$ with support $(\mathcal{A}^l,\mathcal{B}^l)$ satisfying conditions (P1) and (P2).
    \begin{itemize}
        \item[] \textbf{if} $(\mathcal{A}^l,\mathcal{B}^l)$ satisfies (P3),
        \item[] \textbf{then} path $\Gamma^l$ is the path space geodesic,
        \item[] \textbf{else} chose any minimum weight cover $C_1\cup D_2$, $C_1\subset A_i$ and $D_2 \subset B_i$ with complements $C_2$ and $D_1$, respectively, having weight $\frac{||C_1||}{||A_i||}+\frac{||D_2||}{||B_i||}<1$. Replace $A_i$ and $B_i$ in $\mathcal{A}^l$ and $\mathcal{B}^l$ by the ordered pairs $(C_1,C_2)$ and $(D_1,D_2)$, respectively, to form a new support $(\mathcal{A}^{l+1},\mathcal{B}^{l+1})$ with associated path $\Gamma^{l+1}$.
    \end{itemize}
\end{itemize}
\end{alg}

Would be reasonable for the reader to ask how can we assure that $(\mathcal{A}^l,\mathcal{B}^l)$ satisfy (P2) with multiple iterations of \textbf{step}, since (P1) is assured by the construction of $(\mathcal{A}^{l+1},\mathcal{B}^{l+1})$. Owen \textit{et al.} assures that condition on the Lemma 3.4 of \cite{bib:9}.

The biggest slice of the algorithm complexity lays on checking if a specific support satisfies (P3). This is solved through an equivalent problem that is called the \textit{Extension Problem}, but its formalization was to extensive for us to specify here, however, its complexity is $O(n^3)$ \cite[Lemma 3.3]{bib:9}. Adding to that, we need to account for the unsuccessful tries of checking (P3) which are at most $n-3$ (for the maximum $n-2$ possible iterations, that correspond to the maximum number of internal edges that are the higher bound for $|\mathcal{A}|$ and $|\mathcal{B}|$, less the iteration where the solution is found), hence the complexity of our algorithm is $O(n^4)$.

The \textit{Geodesic distance} and the way it's formalized brings a new dimension to visualization. Since there's a continuous path between any two trees, one can technically visualize how the tree deforms along that path. Regarding its \textbf{discriminatory power} it can be seen as holding emphasis according to shared internal edges and its length's between the trees, rather than internal path length (such as the case with \textit{Quartet based metrics}). One could also see it as closer to \textit{Robinson Foulds Length} since each internal edge has a one-on-one correspondence with clades: travessing $\mathcal{T}_n$ can be boiled down to contractions and decontractions of Bourque ($\alpha$ and $\alpha^{-1}$ operations, respectively). To which degree these two metrics relate, is a topic probably worth discussing.

%%%%%%%%%%%%%%%%%%%%%%%%%%%%%%%%%%%%%%%%%%%%%%%%%%%%%%%%%%%%%%%%%%%%%%%%
%%%%%%%%%%%%%%%%%%%%%%%%%%%%%%%%%%%%%%%%%%%%%%%%%%%%%%%%%%%%%%%%%%%%%%%%
%    Other Approaches to the problem                                   %
%         Less usual distances used to approach the problem            %
%           - Maximum Agreement Subtree, Align                         %
%           - Cophenetic Relation, Node, Similarity Based on Probability
%           - Hybridization Number                                     %
%           - Subtree Prune and Regraft                                %
%%%%%%%%%%%%%%%%%%%%%%%%%%%%%%%%%%%%%%%%%%%%%%%%%%%%%%%%%%%%%%%%%%%%%%%%

\section{Other Approches}
\label{section:oapproach}

We will now cover some other metrics that are less used. Some of them might be interesting or promising approaches to the problem at hand, but by some reason are disregarded or unused by the scienctific comunity, such as really expensive computations (as the case of the \textit{Hybridization Number}). We will not, however, cover these distances with the same detail as the ones in the previous section.

\subsection{Maximum Agreement Subtree, Align}
\label{subsection:mast}

The \textbf{Maximum Agreement Subtree} is a concept initially formalized in 1980 by Gordon A.D., and it was covered in \cite{bib:1} when its practical performance was analysed.

\begin{dfn} (\textbf{Maximum Agreement Subtree distance})\\
Let $A,B\in\gamma_S$, $|S|=n$, $T$ the maximum subtree of $A$ and $B$ and $t$ the number of leaves of $T$. So, the Maximum Agreement Subree Distance $$d_{MAST}(A,B)=n-t$$
\end{dfn}

\begin{figure}[!htb]
  \centering
  \includegraphics[width=0.8\textwidth]{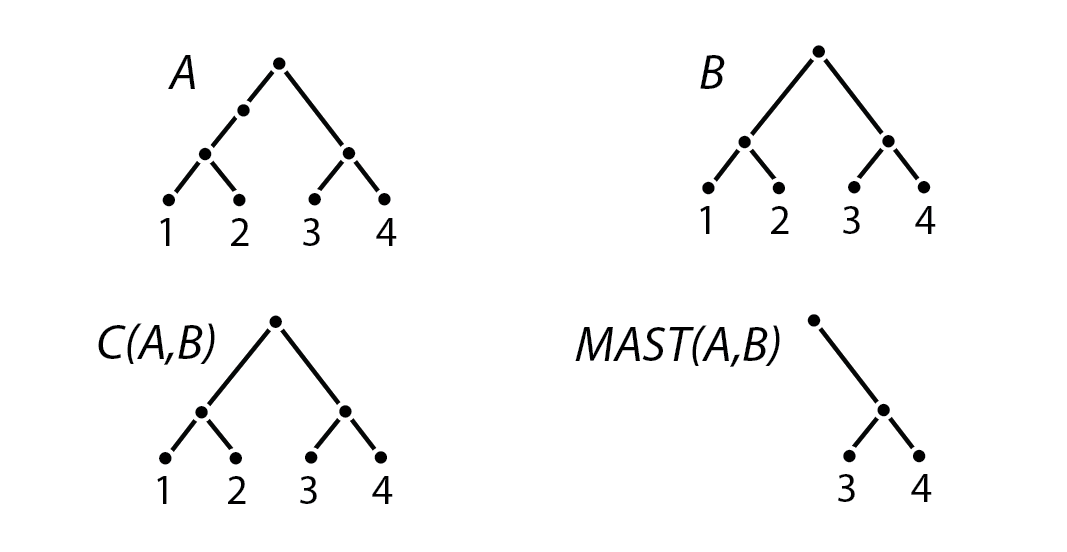}
  \caption[Caption for figure in TOC.]{A simple counterexample to illustrate the difference between a \textit{Maximum Agreement Subtree} and a \textit{Strict Consensus Tree} }
  \label{fig:MAST}
\end{figure}

We could go further on this definition by going through the specifics of what is a ``subtree" (establishing an isomorphism between subsets of $V_A$, $V_B$ and $E_A$, $E_B$ that preserves labeling). One could also be lead to believe that a relation between the \textit{Maximum Agreement Subtree} and the \textit{Strict Consensus Tree}, however, we provide in Figure \ref{fig:MAST} a simple counterexample.

Regarding complexity, the article \cite{bib:18} provides a study of the complexity of computing \textit{MAST}, which the conclusion is that can be solved in $O(n^{O(d)})$ time where $n$ stands for the size of $S$ and $d$ the maximum vertex degree in both trees, concluding that this is a polynomial time problem. However, other papers \cite{bib:1,bib:19} adopt other implementations with other complexities associated: as a matter of fact, imposing some constraints in the data structures can also lead to $O(nlog(n))$ complexity \cite{bib:19}.

Regarding its \textbf{discriminatory power}, based on its definition one would assume that this metric is very sensible to small variations of the same tree, since it only considers what's strictly equal to both trees and disregards everything else. So it should be safe to assume that its power lies close to the one of \textit{Robinson-Foulds}, and should be considered in cases where one desires to analyse how identical to trees are, instead of how similar.

\vspace{10mm}

The first time \textbf{Align} distance was formalized was in 2006 by Nye, \textit{et al.} in \cite{bib:8}. The original motivation behind it was to build an alignment between the two trees at hand (analogous to sequence alignment) building a match between edges according to their topological characteristic. The Align distance is a topologic mesure.

As seen previously, in every tree $A\in\gamma_S$ there is an associated partitioning function (Definition \ref{dfn:partfunc}) that delineates partitions of the set $S$ from the set of edges of $A$, $E_A$. Consider $A,B\in\gamma_S$, $f_A$ and $f_B$ the respective partitioning functions, $e_A\in E_A$, $e_B\in E_B$. Consider as well that, for every $X\in\gamma_S$, partitioning function $f_X$ and edge $e_X\in E_X$, $f_X(e_X)=\{P^X_{(e_X,0)},P^X_{(e_X,1)}\}$. We now define parameters $a_{(r,s)}$ as $$ a_{(r,s)} = \frac{|P^A_{(e_A,r)}\cap P^B_{(e_B,s)}|}{|P^A_{(e_A,r)}\cup P^B_{(e_B,s)}|} $$ that represents the proportion of elements shared by $P^A_{(e_A,r)}$ and $P^B_{(e_B,s)}$. The score $s(e_A,e_B)$ is then defined by $$s(e_A,e_B)=max\{min\{a_{(0,0)},a_{(1,1)}\},min\{a_{(0,1)},a_{(1,0)}\}\}$$

The Align distance is then defined by:

\begin{dfn} \textbf{(Align distance)}\\
Let $S$ be a set of labels of size $n$, $A,B\in\gamma_S$ such that $|E_A|=|E_B|$. The \textbf{Align distance} is given by $$d_{Align}(A,B)=\sum_{e_A\in E_A} s(e_A,f(e_A))$$ where $f:E_A\longrightarrow E_B$ is a bijective function that maximizes the sum. 
\end{dfn}

Finding the bijection $f$ is actually what bounds the time complexity of the implementation, the Munkres Algorithm has $O(n^3)$ time complexity (Munkres, J. \cite{bib:20}; Bourgeois \textit{et al.}\cite{bib:21}).

There is, as stated in \cite{bib:1}, a strong emphasis of the shared clades on the distance function, making it related to \textit{Robinson-Foulds}, when issuing its \textbf{discriminatory power}. But since it uses information from clades that are almost the same the same way it uses from the ones that are the same, the conclusion is that it lies on the opposite side of \textit{MAST} when compared to \textit{RF}.

\subsection{Cophenetic correlation coefficient, Node, Similarity based on Probability}
\label{subsection:coph}

The \textbf{Cophenetic Correlation Coefficient} is known as ``the first effective numerical method known" by most \cite{bib:7}, was originally described by Sokal and Rohlf in 1962 in \cite{bib:22} and its motivation was to measure how faithfull a dendrogram preserves the pairwise distances between the original unmodeled data points. Given the architecture of the method, one can also use it to measure how close the trees at hand are regarding pairwise distance between their leaves.

Another interesting fact is that on \cite{bib:22}, Sokal and Rohlf also stated that \textit{``One of the initial schemes which occurred to us (...) was to try to compare different dendrograms with the same set of leaves by counting the number of breaks and rearrangements necessary to convert one dendrogram into another"} (adapted) and that this would later inspire the original idea of the \textit{Robinson Foulds distance}.

\begin{figure}[!htb]
  \centering
  \includegraphics[width=1\textwidth]{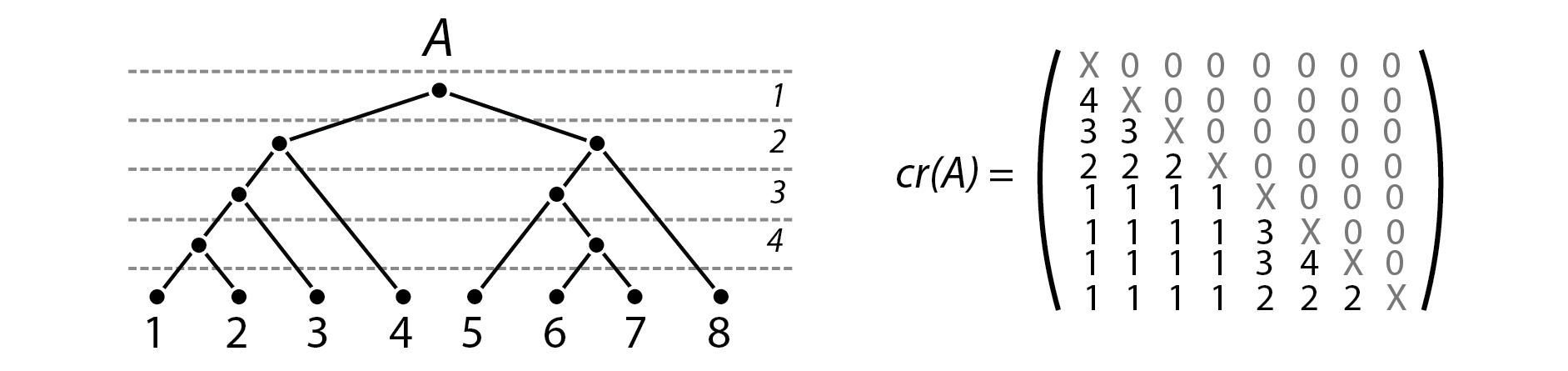}
  \caption[Caption for figure in TOC.]{A dendrogram with internal vertices split for class values and its respective cophenetic relations. It could be the case that vertices with different depths are in the same class value.}
  \label{fig:CCC}
\end{figure}

The first step to calcule the \textit{cophenetic correlation coefficient} for a dendrogram and a corresponding set of data is dividing the internal nodes into suitable \textbf{class values}. These are distributed across the depth of the tree, such that if one node is deeper or as deep as another one, its class value will be greater or equal than the class value of the compared vertex (check Figure \ref{fig:CCC} as an example). 

This process is left to the scientist, keeping in mind that the number of class of values must be picked taking in consideration the number of leaves of the dendrogram (for less than 10 leaves one should stick with 4 or less class values, for 100 leaves one should choose at least 10 class values \cite{bib:22}). Then, the cophenetic relation between two leaves is the class value of the least deep vertex on the path from one leaf to another. The cophenetic relation for a dendrogram is a matrix of size $n \times n$ (where $n$ is the size of the data sample) that comprises all cophenetic relations between its set of leaves. We will denote, for a dendrogram $A\in\gamma_S$, the cophenetic relation of $A$ as $cr(A)$, and the cophenetic relation between leaves $v_i, v_j\in V_A$ is stored on the entry $cr(A)_{ij}$.

\begin{dfn}\textbf{Cophenetic correlation coefficient}
Let $S=\{X_i\}_{i\in [n]}$ a set of data, $\bar{d}$ a distance function for the data, and $A\in\gamma_S$ a dendrogram estimated from the data. The \textbf{cophenetic correlation coefficient} is given by

$$ c(A) = \frac
{\sum^n_{j=1}\sum^j_{i=1} (cr(A)_{ij}-\bar{cr}_A)(\bar{d}(X_i,X_j)-\bar{X})}
{\sqrt{[\sum^n_{j=1}\sum^j_{i=1} (cr(A)_{ij}-\bar{cr}_A)^2][\sum^n_{j=1}\sum^j_{i=1} (\bar{d}(X_i,X_j)-\bar{X})^2]}}
$$

where $\bar{X}$ is the average distance for all possible data pairs from $\{X_i\}_{i\in[n]}$ and $\bar{cr}_A$ the average cophenetic relation between all pairs of vertices for the dendrogram $A$.\\
If we instead want to calculate the cophenetic correlation coefficient between two dendrograms $A,B\in\gamma_S$ we should instead consider

$$ d_{CCC}(A,B) = \frac
{\sum^n_{j=1}\sum^j_{i=1} (cr(A)_{ij}-\bar{cr}_A)(cr(B)_{ij}-\bar{cr}_B)}
{\sqrt{[\sum^n_{j=1}\sum^j_{i=1} (cr(A)_{ij}-\bar{cr}_A)^2][\sum^n_{j=1}\sum^j_{i=1} (cr(B)_{ij}-\bar{cr}_B)^2]}}
$$

where $\bar{cr}_A$ and $\bar{cr}_B$ are the average cophenetic relation between all pairs of vertices for the dendrogram $A$ and $B$ respectively.

\end{dfn}

This coefficient is nothing less than the \textit{Pearson correlation coefficient}, in the latter case of the definition (for two matrices) also referred as the product-moment correlation coefficient between $A$ and $B$.

%%%Regarding its \textbf{complexity}, one should note that after the computations done to obtain $cr(A)$, nothing more is left other than make a calculation that can be done in linear time. \sout{Each entry of $cr(A)$ can be calculated throught Dijktra shortest path algorithm (which's complexity is $O(nlog(n)+e)$) which makes calculating $cr(A)$ a task of $O(n^3log(n))$ time complexity, however, we believe that a more reasonable approach would lead to better results (saving in computations for leaves in distinct clades - Figure \ref{fig:CCC}). }(A ideia que tenho é que isto provavelmente é $log(n)$, basta considerar as partições de $S$ e respectivos subconjuntos a partir da remoção das raizes de cada uma das clades do grafo. Ou seja: a partição de $S$ $\{\{v_1,...,v_i,...,v_k\},\{v_{k+1},...,u_j,...,u_n\}\}$ determinada por ao remover a raiz de uma árvore $A\in\gamma_S$ devolve sempre dois conjuntos tal que, para todo $cr(A)_{ij}=1$. Assim constrói-se $cr(A)$ recursivamente).

Regarding its \textbf{complexity}, one should note that after the computations done to obtain $cr(A)$, nothing more is left other than make a calculation that can be done in linear time. To calcule every entry of $cr(A)$ we can do a calculation that will take $O(log(n))$ time. Every vertex $v$ of our dendrogram $A$ induces a partition of its labels dictated by the leaves of the trees in the forest generated by removing $v$ from $A$. The cophenetic relations between the leaves on these partitions is given by the class value of the vertex $v$ removed. From this point on we can apply this method recursively to the neighborhood of the removed vertex $v$ and by the end of the computation $cr(A)$ will be calculated.

Considering the tree $A$ in the Figure \ref{fig:CCC}, removing the root vertex would give us the partition of $S$ $\{P^1_1,P^1_2\}=\{\{1,2,3,4\},\{5,6,7,8\}\}$, concluding that $cr(A)_{ij}$ equals the class value of the root vertex, which is $1$, for all $i\in P^1_1$ and $j\in P^1_2$. From this point on, we would apply the same procedure at the child vertices of the root vertex (that lay in class value $2$), which would give partitions $\{\{1,2,3\},\{4\}\}$ and $\{\{5,6,7\},\{8\}\}$.

This was, as stated, the first effective numerical method to compare classifications, there are limitations and concerns as far its \textbf{discriminative power}. Williams \textit{et al.}, in its article about a variant of this metric which we'll approach next, refers how this metric has drawbacks in regards of how the class values are not a property from the classification itself, but something instead defined by the scientist \cite{bib:7}. This alone makes the \textit{Cophenetic correlation coeficient} is obsolete.

\vspace{10mm}

The \textbf{Node distance} was formalized by W. T. Williams and H. T. Clifford in 1971 \cite{bib:7} and is a variant of the cophenetic correlation coefficient pairwise heuristic, looking forward to improve on the limitation in discriminatory power (as pointed out in \cite{bib:7}).

\begin{dfn} \textbf{(Node distance)}\\
Let $S$ be a set of labels such that $|S|=n$, $A,B\in\gamma_S$ and, for all tree $X\in\gamma_S$ and $v_i,v_j\in V_X$, $d^X(v_i,v_j)$ the distance of the path from $v_i$ to $v_j$ in the tree $X$. Consider as well a function $l_A$ and $l_B$ that, given a label $s\in S$, $l_A(s)$ and $l_B(s)$ returns the vertices with label $s$ from $V_A$ and $V_B$ respectively (the $label$ ``inverse" function). The \textit{Node distance} function is given by

$$d_N(A,B)=\frac{2\sum_{s_1,s_2\in S}|d^A(l_A(s_1),l_A(s_2))-d^B(l_B(s_1),l_B(s_2))|^k}{n(n-1)}, k=1 $$

\end{dfn}

Article \cite{bib:1} also refers that a similar metric as proposed by Penny, et all in 1982 which follows the \textit{Node distance} outlines but for $k=2$ instead, by the name of \textit{Path Difference metric}.

The \textbf{complexity} of the \textit{Node distance}, just as the \textit{Cophenetic Correlation Distance}, is determined before the actual calculation, this time by determining the distance between every leaf of the two subject trees. The path length can be obtained through a topological sorting algorithm, of complexity $\Theta (|V|+|E|)$ for some tree $T=(V,E)$. Since this operation will be done once for each labeled vertex, the time complexity of \textit{Node} will be $O(n^2)$.

As for the \textbf{discriminatory power} goes, Penny, \textit{et al.} refers to \textit{Path Difference metric} in \cite{bib:26}: describes it as ``sensitive to the tree distribution" (since its formulation wasn't accounting for $n(n-1)$ division, which is done here) and points out that another usefull application would be ``when the topic of interest is the relative position of subsets of nodes, rather than the comparisson of trees themselves" (adapted).

\vspace{10mm}

The \textbf{Similarity based on probability} is a metric firstly defined in \cite{bib:23} by Hein \textit{et al.} and like most metrics defined in this work, it was studied in \cite{bib:1}. This metric is rather unique compared to the others given its probabilistic approach, which suits the book where its integrated. The following definition was adapted from Chapter 7 of \cite{bib:23}.

\begin{dfn} \textbf{(Similarity based on probability)}\\
We start by defining an indicator function $I$ and measure $M$, where $r_X\in V_X$ is a vertex such that $label_X(r_X)='root'$ and $a^X_u\in V_X$ is the ancestor of $u\in V_X$ in the tree $X$ 
$$I_{\{u,v\}}=
\begin{cases}
1 & if\ f_A(u)=f_B(v)\\
0 & otherwise
\end{cases};$$\\$$
M_{XY}=\frac{\sum_{u\in V_X\backslash\{r_X\};v\in V_Y\backslash\{r_Y\}}I_{\{u=v\}}w(a^X_u u)w(a^Y_v v)}{l_X l_Y}
$$
For all $X,Y\in\gamma^w_S$ (with $f_X$ and $f_Y$ as the respective partitioning functions - Definition \ref{dfn:partfunc} ), $u\in V_X$, $v\in V_Y$, $l_X=\sum_{e\in E_X} w(e)$ and $l_Y=\sum_{e\in E_Y} w(e)$. We now define the \textbf{similarity based on probability} function $S$, for two weighted trees $A,B\in\gamma^w_S$, as

$$S(A,B)=\frac{M_{AB}}{M_{AA}}$$
\end{dfn}

The meaning behind this similarity measure $M_{AB}$ is, as described in \cite{bib:1}, the \textit{``probability that a point chosen randomly in A will be on a branch leading to the same set of tips as a point chosen randomly in B"}, which is afterwards normalized by $M_{AA}$. This leads to a non-symmetry scenario (a requirement for a distance), which is solved in \cite{bib:1} in the following way:

\begin{dfn}\textbf{(Similarity based on probability distance)}\\
Let $A,B\in\gamma^w_S$ and $S$ the similarity based on probability function. We define the \textbf{Similarity based on probability distance} $d_{Sim}$ as
$$d_{Sim}(A,B)=1-\frac{S(A,B)+S(B,A)}{2}$$
\end{dfn}

The \textbf{time complexity} of this algorithm lies on the indicator function, since every other step are merely calculations which can be done in linear time. To obtain the partitioning described the the functions $f_X$ and $f_Y$ one should apply a breadth-first search which complexity is $O(|V|+|E|)$, and since this must be done for every for every vertex on each tree (storing results not to repeat calculations), the time complexity of the indicator function is $O(|V|^2)$.

Regarding its \textbf{discriminatory power}, article \cite{bib:1} reported that its performance was underwhelming for trees with five tips and with branch length zero, but excluding these cases behaved similarly to other branch length metrics. It also states that for problems where branch proportion are important but their absolute value isn't, this should be the selected metric.

\subsection{Hybridization Number}
\label{subsection:hyb}

This fairly recent concept was brought up around mid of the first decade of the two thousands, and the main concept behind it is the assumption that evolution doesn't need to be described by a tree structure, since cross-breeding can be an event behind a species existence. Cross-breeds are often called \textit{hybrids}, hence the name of this concept. 

That lead us to expand the standard data structure we've worked until this point, since now we can have two distinct paths from the root to a leaf. Following the main motivation for this concept, \textit{directed acyclic graphs} (or DAGs) suit discussed problem. The article followed in our research was \cite{bib:6}, since it gives a good introduction to the subject, even if its main purpose is presenting results regarding the complexity of the problem that we will approach right after defining some key concepts. Since in \textbf{directed graphs} the edges $(a,b)$ and $(b,a)$ are different, we need to adapt some concepts such as the degree of a vertex:

\begin{dfn} \textbf{(In and Out-degree, Split and Reticulation Nodes)}\\
Let $T=(V,E)$ be a directed graph (a graph where, for $u,v\in V$, the edges $uv$ and $vu$ are different). The \textbf{in-degree} of a vertex $u$, denoted as $d^-(u)$, is determined by $|E_{\cdot u}|$ where $E_{\cdot u}=\{xu\in E: x\in V\}$. Similarly, the \textbf{out-degree} of $u$, denoted as $d^+(u)$, is determined by $|E_{u\cdot}|$ where $E_{u\cdot}=\{ux\in E: x\in V\}$.\\
A \textbf{split node} of a directed graph $T=(V,E)$ is a $u\in V$ such that its in-degree is at most 1 and its out-degree at least 2. A \textbf{reticulation node} of a directed graph $T$ is a $v\in V$ such that its in-degree at least 2 (reticulations for short).
\end{dfn}

\begin{dfn} \textbf{(Hybridization Number)}\\
For a graph $T=(V,E)$, the \textbf{hybridization number} $H_T$ is given by $$H_T=\sum_{v\in V} (d^-(v)-1)$$
\end{dfn}

\begin{dfn} \textbf{(Hybridization Problem; Hybridization Distance)}\\
 Given a forest $\mathcal{F}=\{T_0,T_1,...,T_k\}$, where $T_i\in\gamma_S$ for every $0\leq i \leq k$, the \textbf{hybridization problem} consists in finding a graph (which we refer to as \textbf{\textit{hybridization network}}) $N$ such that:
\begin{itemize}
 \item[] $(1)$ For every $T_i$ there is an injective map $h_i:V_{T_i}\rightarrow V_N$ that preserves vertex adjacency (that is, if $uv\in E_{T_i}$ then $h_i(u)h_i(y)\in E_N$);
 \item[] $(2)$ $H_N$ is minimum amoungst all trees that satisfy $(1)$.
\end{itemize}

Assuming this problem is solved by $\mathcal{P}$, we can now define a distance $d_H$ between two trees $A,B\in\gamma_S$ as $$d_H(A,B)=H_{\mathcal{P}(\{A,B\})}$$ were $\{A,B\}$ is the forest for $\mathcal{P}$. We name this distance the \textbf{hybridization distance}.
\end{dfn}

As stated in \cite{bib:6}, \textit{``The holy grail for this problem is to develop algorithms that can cope with many input trees and non-binary input trees"}, since there's no actual efficient way to compute such metric. We are talking about a metric in which research is still beeing done given its good interpretation value on phylogenetics, and even though it's beeing formalized for input sets with arbitrary number of trees, computing the problem for two specific trees it's a problem considered to lay in \textbf{NP-Hard} and \textbf{APX-Hard}.

As far as its \textbf{discriminatory power} goes it's interesting to note that this metric values not only the shared clades between two trees, but the clades in which its cluster representation are not disjoint. Also, the interpretative value for philogeny is fairly relevant in this case, however, it's still relatively early to come with practical conclusions regarding its discriminatory power since testing is not yet a viable task.

\subsection{Subtree Prune and Regraft}
\label{subsection:SPR}

The idea of \textbf{Subtree Prune and Regraft} distance goes back to Sokal and Rohfl idea of identifying how many operations are two trees apart from eachother. This operation, which is named \textit{prune and regraft}, has a far more relevant interpretation in phylogeny when compared to the $\alpha$ operation from where Robinson and Foulds started drafting, and actually, that's the main reason behind the intense research and study made around this operation. As for now (and just like the hybridization number), only looks promising since computing this operation is far from a trivial task from a complexity standpoint.

Our main resource for this subject was \cite{bib:2}, a paper on optimization of the \textit{Subtree Prune and Regraft}, or \textit{SPR}, from 2016. This is a very complete article that compiles not only a good background for understanding the intricasies of the subject at hand, as a handfull of important results towards a practical and usable distance formalization. As for us now, lets us go over the definition of the \textit{SPR} operation and remarkable results.

\begin{dfn} \textbf{(Rooted Subtree Prune and Regraft operation - rSPR)}\\
Let $A\in\gamma_S$, $\rho\in V_A$ the root of $A$, $u\in V_A\backslash\{\rho\}$, $E_{C_u}\subseteq E_A$ the set of edges of the clade defined by $u$, the vertex $v\in V_A$ as the ancestor of $u$, $Adj_v=\{e_1,e_2,...,e_k\}\subseteq E_A$ the set of edges connecting $v$ to its neighboors, and $xy\in E_A\backslash (E_{C_u}\cup Adj_v)$. The \textbf{rooted Subtree Prune and Regraft} operation is a function $uSPR:\gamma_S\times V\times E\longrightarrow \gamma_S$ such that $uSPR(A,u,xy)=(V,E,S)$ generated by the following procedure:

\begin{itemize}
  \item[] $(1)$ $V_0=V_A\cup\{v'\}$; $E_0=(E_A\backslash\{vu,xy\})\cup\{xv',v'y,v'u\}$;
  \item[] $(2)$ Remove $vu$ from $Adj_v$ and relabel its elements;
  \item[] $(3)$ $(V,E,S)=\alpha(\alpha(...\alpha((V_0,E_0,S),e_{k-1})...,e_2),e_1)$;
  \item[] $(4)$ if $v'$ is adjacent to $\rho$ then $label(\rho)='NULL'$ and $label(v')='root'$.
\end{itemize}
\end{dfn}

The \textbf{unrooted Subtree Prune and Regraft} is also defined as the previous operation in data structures without root, leaving aside all requirements and steps that involving it, and also disregarding the requirement that $v$ is an ancestor of $u$.

\begin{dfn} \textbf{(Subtree Prune and Regraft distance)}\\
  Let $A,B\in\gamma_S$. The Subtree Prune and Regraft distance, denoted by $d_{SPR}$, equals the number of SPR operations required to transform $A$ in $B$.
\end{dfn}

\begin{nota}
 Let $A\in\gamma_S$ and $S'\subseteq S$. $A(S')$ is the minimal rooted subtree of $A$ that connects the leaves labeled with $S'$. Furthermore, we denote by $A|S'$ the tree generated from $A(S')$ with every non-root vertex of degree 2 supressed.
\end{nota}

A lot of work has been put recently on researching about this operation. Allen and Steel (2001) proved a theorem about a distance defined the same way as this previous one, but for a more general operation that relates to SPR, the \textit{tree-bisection-reconnection} (or TBR for short) \cite{bib:24}. Ultimately, that lead to Bordewich and Semple (2005) prooving the same conclusion in regards of our SPR operation \cite{bib:25}. We'll expose that result after exposing the concept of \textit{maximum agreement forest}. This was stablished for \textit{rooted binary trees}, but for other data structures should work similarly:

\begin{dfn} \textbf{(Maximum Agreement Forest)}\\
Let $A,B\in\gamma_S$ be binary rooted trees. An \textbf{agreement forest} for $A$ and $B$ is a collection $\mathcal{F}=\{T_\rho,T_1,T_2,...,T_k\}$ where $T_\rho\in\gamma_{S_\rho},T_1\in\gamma_{S_1},T_2\in\gamma_{S_2},...,T_k\in\gamma_{S_k}$ and $T_1,T_2,...,T_k$ are binary and the following properties are satisfied:
\begin{itemize}
  \item[] $(1)$ $S_\rho,S_1,S_2,...,S_k$ partition $S\cup\{\rho\}$ and, in particular, $\rho\in S_\rho$;
  \item[] $(2)$ For every $i\in\{\rho,1,2,...,k\}$, there is bijective maps that preserve labeling between $T_i$, $A|S_i$ and $B|S_i$;
  \item[] $(3)$ The trees in $\{A(S_i):i\in\{\rho,1,2,...,k\}\}$ and $\{B(S_i):i\in\{\rho,1,2,...,k\}\}$ are vertex-disjoint subtrees (trees which their vertex set are disjoint) of $A$ and $B$, respectively.
\end{itemize}

The \textit{agreement forest} in which the $k$ is minimised is called \textbf{maximum agreement forest} and that $k$ is denoted by $m(A,B)$.
\end{dfn}

For unrooted trees, the previous definition holds without the requirement of $S_\rho$ and $T_{S_\rho}$.

\begin{thm} (Bordewich and Semple, 2005)\\
Let $A,B\in\gamma_S$. Then, the \textbf{Subtree Prune and Regraft distance} is given by
$$d_{SPR}(A,B)=m(A,B)-1$$
\end{thm}

Regarding \textbf{complexity}, there is, until the time of publication of \cite{bib:2}, that beeing 2015, no solid idea on the complexity of the \textit{SPR} distance, however, it's stated that is conjectured that another distance, by the name of \textit{replug}, that captures a lot of \textit{SPR} distance properties, is \textbf{NP-Hard}, which leads to believe that \textit{SPR} falls under the same category. However, and how we stated before, this is still an ongoing topic of discussion due to its application relevance.

It's interesting to understand, given the metric formulation, how \textit{SPR} shares some of \textit{Robinson Foulds} characteristics regarding its \textbf{discriminatory power}, but it's also interesing to try to understand how promising it might be, given that the limitation of small variations on the trees from the \textit{RF} distance is solved with a simple vertex replug. Just like the \textit{Hybridization Number}, the interpretative value for philogeny is fairly relevant in this case, however, it's still relatively early to come with practical conclusions regarding its discriminatory power since testing is not yet a viable task.

%%%%%%%%%%%%%%%%%%%%%%%%%%%%%%%%%%%%%%%%%%%%%%%%%%%%%%%%%%%%%%%%%%%%%%%%

\section{Distance synopsis}
\label{section:syn}

After all the formalization and property lifting of all the metrics in previous sections, we now present two tables compiling the information regarding the complexity and discriminative power of all the metrics. Since $d_{H}$ and $d_{SPR}$ don't have yet feasible algorithms to be calculed, it's presented in Table \ref{tab:dprob} the conjectured problem hardness of its computation.

\begin{table}[!ht]
\centering

\label{tab:dcomp}
\begin{tabular}{|l|l|l|}
\hline
            & Time Complexity                         & Discriminatory Power                                                                                                                                                                                                                                       \\ \hline
$d_{RF}$    & $O(n)$                                 & \begin{tabular}[c]{@{}l@{}}- Really sensible to the scalability of S; \\ - Unstable: moving a single leaf could lead\\ to great discrepancies in the distance\\ value; \\ - Overperforms in close to resolved trees\\ other than to unresolved ones.\end{tabular} \\ \hline
$d_{RFL}$   & $O(n)^{(*)}$                           & \begin{tabular}[c]{@{}l@{}}- Shares discriminatory power from RF; \\ - In some cases is non-symmetric and\\ doesn't has identity of indiscernibles.\end{tabular}                                                                                              \\ \hline
$d_Q$       & $O(dnlog(n))$ to $O(n)$                 & \multirow{3}{*}{\begin{tabular}[c]{@{}l@{}} - Monotonous with the scalability of S; \\ - More sensible to alterations in the\\ bottommost branches; \end{tabular}}                    \\ \cline{1-2}
$d_{Trip}$  & \multirow{2}{*}{$O(nlog(n))$ to $O(n)$} &                                                                                                                                                                                                                                                            \\ \cline{1-1}
$d_{TripL}$ &                                         &                                                                                                                                                                                                                                                            \\ \hline
$d_{Geo}$   & $O(n^4)$                                & \begin{tabular}[c]{@{}l@{}} - Emphasise shared internal edges and its\\ lengths rather than internal path lengths; \\ - Close to RF discriminatory power.\end{tabular}                                                                                        \\ \hline
$d_{MAST}$  & $O(|S|^{O(d)})$                         & \begin{tabular}[c]{@{}l@{}} - Very sensible to small variations; \\ - Should be used to analyse how identical\\ two trees are rather than similar; \\ - Close to RF discriminatory power.\end{tabular}                                                           \\ \hline
$d_{Align}$ & $O(n^3)$                                & \begin{tabular}[c]{@{}l@{}} - Strong emphasis in shared clades; \\ - Related to RF discriminatory power, but\\ not close: it values clades that are similar\\ but not identical.\end{tabular}                                                                    \\ \hline
$d_{CCC}$   & $O(n)$                                  & \begin{tabular}[c]{@{}l@{}} - First effective numerical method to\\ compare classifications; \\ - Subjective parameters make this metric \\not so precise and obsolete.\end{tabular}                                                                             \\ \hline
$d_{N}$  & $O(n^2)$                                & \begin{tabular}[c]{@{}l@{}} - Sensitive to tree distribution {[}26{]}; \\ - Better used when the relative position of\\ subsets of nodes is more important than\\ actual tree comparison {[}26{]}.\end{tabular}                                                  \\ \hline
$d_{Sim}$   & $O(|V|+|E|)$                            & \begin{tabular}[c]{@{}l@{}} - Excluding some particular cases, it\\ behaves similarly to other branch length\\ metrics {[}1{]}; \\ - Good metric to use when branch weight\\ proportion is important despise their\\ absolute value {[}1{]}.\end{tabular}         \\ \hline
\end{tabular}
\caption{Synopsis and comparisson of distance's discriminatory power with defined time complexity. In this table, $d$ stands for the maximum vertex degree in the trees in which the distance is beeing calculated. In $(*)$ the denoted complexity is for trees in which the mapping function $h_{(A,B)}$ is unique, otherwise the complexity is undefined.}
\end{table}

\pagebreak

\begin{table}[!ht]
\centering
\begin{tabular}{|l|l|l|}
\hline
          & Problem Hardness & Discriminatory Power                                                                                                                                                                                                                                       \\ \hline
$d_{Hyb}$ & \begin{tabular}[c]{@{}l@{}} APX-HARD,\\ NP-HARD \end{tabular}         & \begin{tabular}[c]{@{}l@{}} - Values not only the shared clades between\\ classifications, but the  clades in which\\ their cluster representation is not disjoint; \\ - Interpretative value for phylogeny sets it\\ apart from the other metrics.\end{tabular} \\ \hline
$d_{SPR}$ & NP-HARD$^{(*)}$               & \begin{tabular}[c]{@{}l@{}} - May solve the RF problems in trees with\\ small variations given the replug move; \\ - Being the interpretative value for\\ phylogeny fairly relevant, $d_{SPR}$ is fairly\\ promising.\end{tabular}                              \\ \hline
\end{tabular}
\caption{Synopsis and comparisson of distance's discriminatory power with undefined time complexity. In $(*)$ the denoted problem hardness is estimated.}
\label{tab:dprob}
\end{table}

\section{Final remarks and challenges}
\label{section:challenges}

The comparisson of classifications, since its genesis in phylogenetics, has come a long way, and its advancements were beneficial in pretty much in everything that benefited from drawing distances between tree-like structures (as an example out of phylogenetics, they could be used as a way to compare algorithms by their running tree). 

However, given the field of study that bloomed all these methods (alongside the fact that was still a research topic), rigour and formalization was not the top-most priority, as we can see from our references. Formalizing or attempting to formalize these metrics (as we just did) is a challenge by itself and even though we see our effort and result as mostly satisfying, there's still some cases where the definition isn't as clean as possible (such as the case of \textit{Quartet based metrics} and defining the space of trees $\mathcal{T}_n$ for every $n\in\mathbb{N}$).

There's the discussed problem with \textit{Robinson Foulds Length} that was exposed in the respective subsection: $d_{RFL}$ is formalized on top of the existence of a matching function $h_{(A,B)}$ which might not be unique. That's a problem since if there exists more than one matching function, the distance between two trees might be undefined given the existance of two possible results. More than that, but this distance is not symmetric as well, meaning that it might be the case that $d_{RFL}(A,B)\neq d_{RFL}(B,A)$. One interesting researching topic might be studying the viability of correcting these problems while mantaining the metric properties and keeping it close to its initial formulation.

When we approached the \textit{Geodesic distance}'s discriminatory power we brought up that $d_{Geo}$ actually relates to $d_{RF}$ since traversing $\mathcal{T}_n$ actually corresponds to contracting and decontracting edges. Even though they are formalized for different structures (the \textit{Geodesic distance} is formalized for weighted trees, contrary to \textit{Robinson Foulds}), if we put this fact aside we can actually examine examples to see more clearly the relation between these two metrics.

\begin{figure}[!htb]
  \centering
  \includegraphics[width=0.8\textwidth]{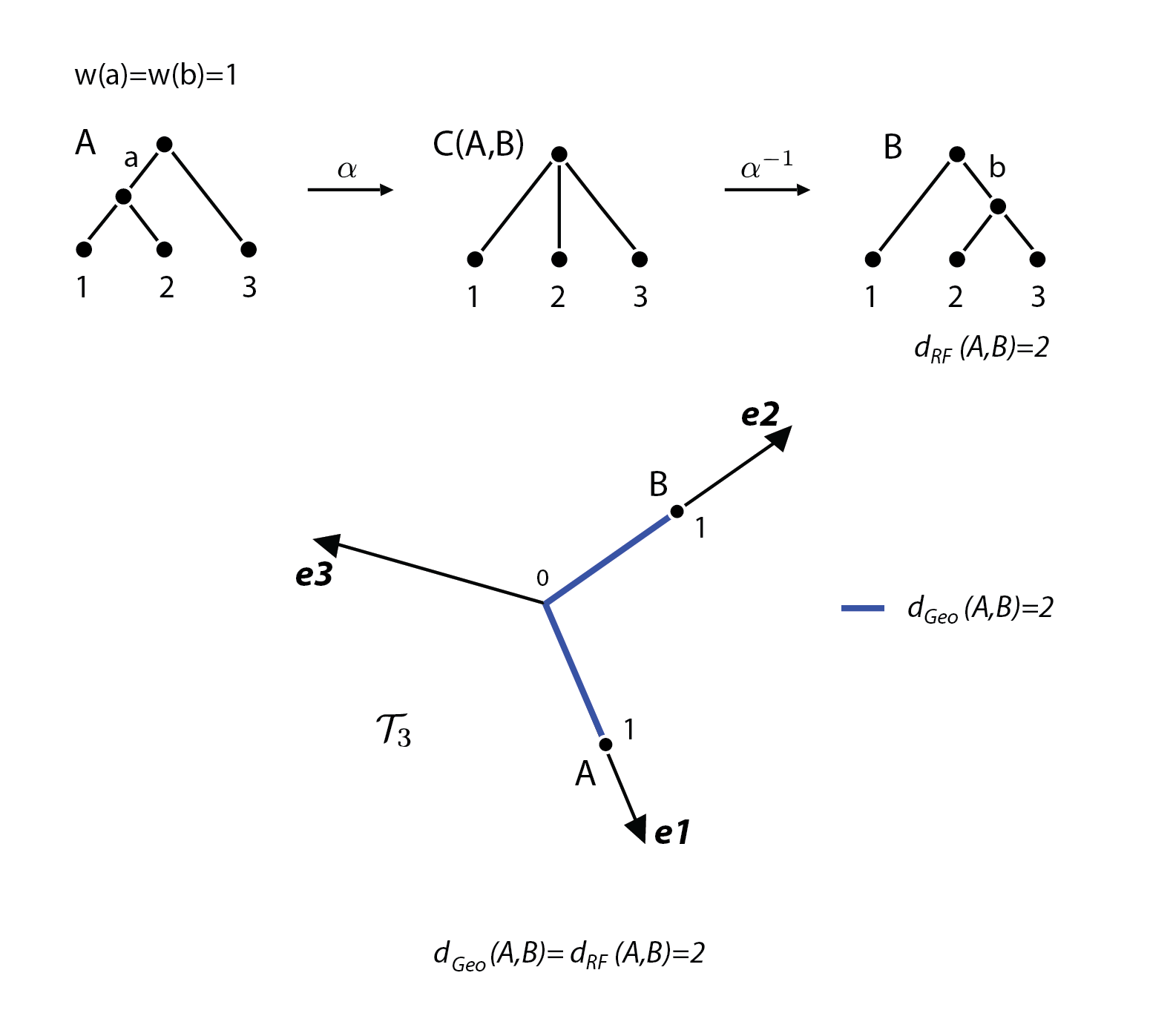}
  \caption[Caption for figure in TOC.]{Visualization of an example where $d_{RF}$ and $d_{Geo}$ coincide.}
  \label{fig:RFGeoEq}
\end{figure}

As we can see in the Figure \ref{fig:RFGeoEq}, the trees $A$ and $B$ are set appart by two $\alpha$ operations, and the geodesic distance between $A$ and $B$ in $\mathcal{T}_3$ is the cone path between these two trees. If we assume the weight of the internal edges $a$ and $b$, the cone path between $A$ and $B$ is indeed $2$.
However, even when we consider the length of the internal edges as $1$, it's not always the case that $d_{Geo}$ and $d_{RF}$ coincide, since contractions and decontractions can happen \textit{simultaneously} while travessing the space of trees.

\begin{figure}[!htb]
  \centering
  \includegraphics[width=0.8\textwidth]{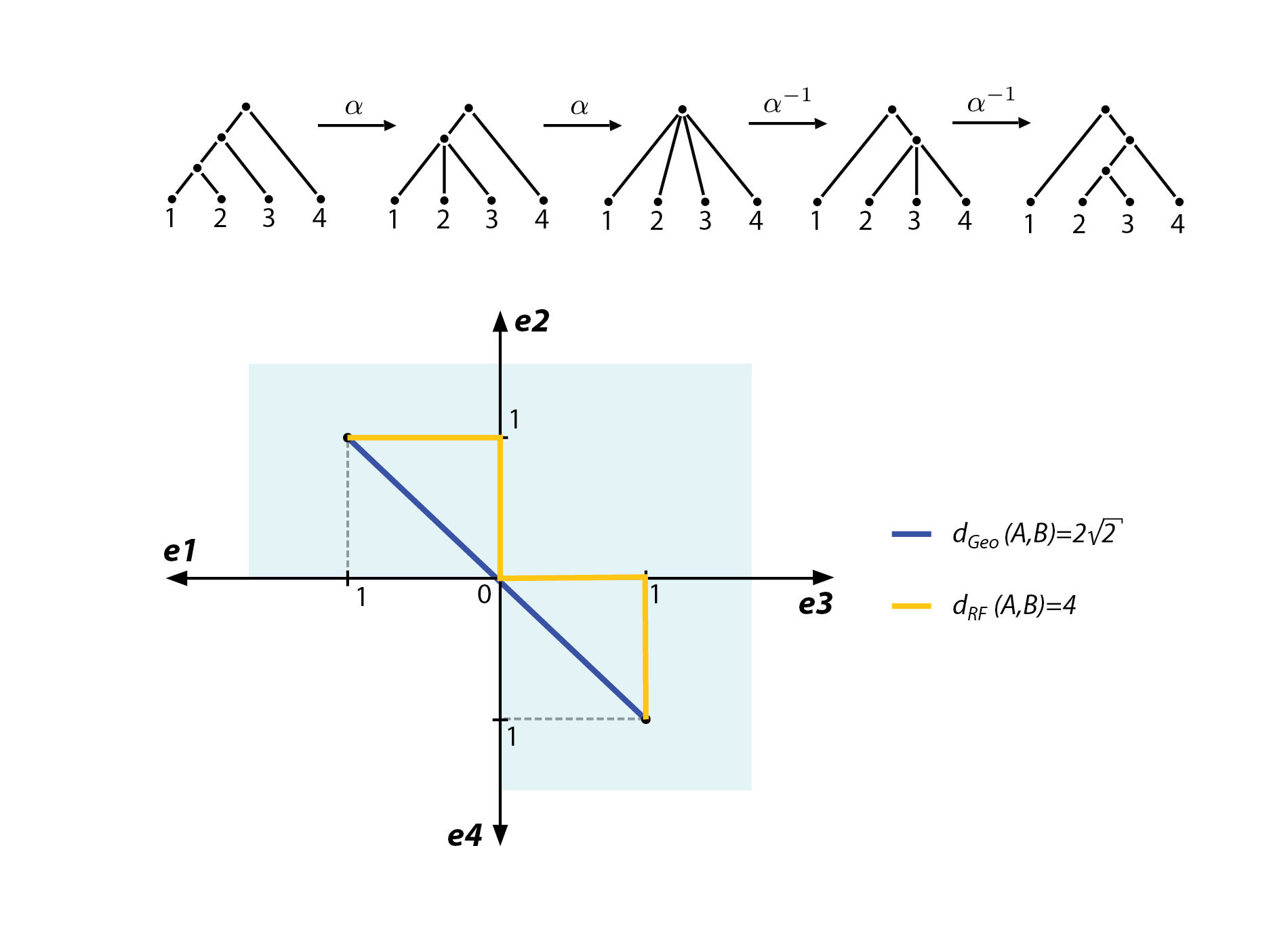}
  \caption[Caption for figure in TOC.]{Visualization of an example where $d_{RF}$ and $d_{Geo}$ don't coincide.}
  \label{fig:RFGeoDif}
\end{figure}

Such is the case in Figure \ref{fig:RFGeoDif}. While in $d_{RF}$ four $\alpha$ operations are done, in $d_{Geo}$ each pair of contractions weight $\sqrt{2}$ for the final result, since they are beeing done ``at the same time".
This relation between these two metrics could be further studied, as a way to strengthen our understanding between their link: understanding the cases in which the metrics coincide might be a good way to improve the computation time of $d_{Geo}$ for big datasets. The same way we could use $d_{RF}$ to optimize the usage of $d_{Geo}$, a study could be made in how could we use more efficient methods to bypass some cases on the less efficient ones. From a practical point of view, these approaches would be an advantage given the big scope of the data handled on the fields of application, nowadays. Also, this approach doesn't need to be exact, from a practical point of view an approximation can often be good enough.

Another thing that might be interesting to explore is the relation between the discriminatory power and the efficiency of the metric: it seems that the less efficient metrics are, greater is their discriminatory power. However, this might also be related to the age of the metrics, since the optimization of oldest distances were also subject of study for longer than the most recent ones. New distances are born from the limitations of the ones that already exist, so it should be no surprise that they are more powerfull from a discriminative standpoint.

\bibliography{refer}
\bibliographystyle{acm}

\nocite{*}

\end{document}